\documentclass[12pt,letterpaper]{article}
\usepackage{amsfonts}
\begin{document}

\def\seCtion#1{\section{#1} \setcounter{equation}{0}}
\renewcommand\theequation{\ifnum\value{section}>0{\thesection.
\arabic{equation}}\fi}

\pagestyle{empty}  
\begin{titlepage}
\begin{flushright}
FIUN-CP-02/1
\end{flushright}
\quad\\
\begin{center}
{\LARGE{\bf Spontaneous CP Phases }}\\
\medskip
{\LARGE{\bf and Flavour Changing Neutral Currents }}\\
\medskip
{\LARGE{\bf in the Left-Right Symmetric Model }}\\
\vspace{1.5cm}
{\bf Yeinzon Rodr\'{\i}guez\footnote{Researcher of the Centro de
Investigaciones of the Universidad Antonio Nari\~{n}o, Cll 58A {\#} 37-94,
Bogot\'a D.C., Colombia.

Email Address: yeinzon@andromeda.uan.edu.co} and Carlos Quimbay\footnote{Associate
Researcher of the Centro Internacional de F\'{\i}sica, Ciudad Universitaria,
Bogot\'a D.C., Colombia.

Email Address: carloqui@ciencias.unal.edu.co}}\\
\vspace{0.2cm}
{\it Departamento de F\'{\i}sica, Universidad Nacional de Colombia\\
Ciudad Universitaria, Bogot\'a D.C.\\
Colombia\\}

\vspace{1.5cm}

\begin{abstract}
We study the behaviour of the flavour changing neutral currents in
the Left-Right Symmetric Model related to the presence of spontaneous
$CP$ phases. To do this, we explore four cases corresponding to
combinations of maximal and no $CP$ violation in both the
lepton and quark sector. We find that we can constrain the flavour
changing neutral currents to the experimental limit, by adjusting the  
$CP$-violating phase of the quark sector, opening
the possibility to obtain a large $CP$ violation in the lepton sector
as well as new Higgs bosons at the electroweak scale.

\vspace{1cm}

PACS Codes: 11.30.Er; 11.30.Hv; 12.60.-i; 12.60.Fr

\vspace{5mm}

{\it Keywords:} $CP$ Violation; Flavour Changing Neutral Currents;
Left-Right Models; Lepton-Sector $CP$ Violation; New Higgs Bosons
\end{abstract}
\end{center}

\end{titlepage}

\newpage
\pagestyle{plain}

\seCtion{Introduction}

The Left-Right Symmetric Model $\left( LRSM\right) $ was developed looking
for a natural origin for parity violation. \cite{leftright} Introducing a
new group of symmetries, $SU\left( 2\right) _R$, we can have a new ``weak''
interaction with the same coupling constant of the weak interaction of the
Standard Model $\left( SM\right) $, but acting only on the right-handed
particles. The resulting model is very interesting because it introduces new
gauge bosons which we can search for in accelerator experiments. The
experimental constraints on the masses of the three new gauge bosons $\left(
W_R^{+},W_R^{-},Z_R^0\right) $ tell us that they have to be at least of $715$
$GeV$ for the charged bosons, and $564$ $GeV$ for the neutral one. \cite{pdg}
If we have only one bidoublet of Higgs bosons in order to improve the Higgs
mechanism in this kind of models with left- and right-handed particles
organized in doublets, we will get the same masses for the left-weak and the
right-weak bosons with equal electric charge. The usual way to avoid this
problem is by introducing new Higgs bosons in the model. The most popular
method to do so in the $LRSM$ is through two Higgs triplets each one acting
on one type of particles depending on its chirality. \cite{langacker}-\cite
{deshpande} With these two triplets, we are able to explain the small masses
of the left-handed neutrinos by the so-called see-saw mechanism, whilst
giving experimentally compatible heavy masses to the right-handed neutrinos.  
\cite{deshpande}-\cite{khasanov}

New interesting things appear: The $SM$ is based on the group of symmetries $%
SU\left( 3\right) _C\otimes SU(2)_L\otimes U(1)_Y$, \cite{weinberg} where
the values that the hypercharge quantum number $Y$ takes for each particle
do not have any physical meaning. When we extend the $SM$ to the $LRSM$, we
have the group of symmetries $SU\left( 3\right) _C\otimes SU(2)_L\otimes
SU\left( 2\right) _R\otimes U(1)_{B-L}$, where the hypercharge quantum
number $Y$ now becomes $B-L$, the difference between the baryonic number $B$
and the leptonic number $L$. \cite{b-l}

Another interesting thing in the $LRSM$ is the possibility of explaining the
observed $CP$ violation. The $SM$ is able to explain it because there are
three families of particles. With three families of particles it is possible
to absorb all the complex phases arising from the Yukawa sector of the
Lagrangian except for one. This remaining phase appears in the
Cabibbo-Kobayashi-Maskawa $\left( CKM\right) $ matrix and it finally can
explain the observed $CP$ violation in the kaon system. \cite{ckm} In the $%
LRSM$ we can impose a global $CP$ symmetry on the complete lagrangian, in
order to avoid explicit complex phases in the Yukawa couplings, and obtain
them spontaneously through the vacuum expectation values arising from the
symmetry breaking mechanism. \cite{barenboim1, barenboim3, spontaneous}-\cite
{spontaneous3} Two spontaneous $CP$ phases appear, $\alpha $ and $\theta $,
which may be allocated in the $CKM$ matrix and in the analogous matrix for
the lepton sector respectively. Once again, this is really interesting
because it opens the possibility of having $CP$ violation in the lepton
sector too, which is not possible in the $SM$.

However, everything is not perfect in the $LRSM$. Due to the presence of a
bidoublet of Higgs fields, Flavour Changing Neutral Currents $\left(
FCNC\right) $ appear in the model involving the neutral scalar bosons. \cite
{barenboim1, deshpande, fcnc, fcnc2} The experimental constraints tell us
that if the $FCNC$ exist, they must be suppressed enough so that the
experiments will not be sensitive to them. \cite{pdg} The $SM$ does not
present any $FCNC $ because it has only one doublet of Higgs fields. One
possible way to avoid the $FCNC$ in the $LRSM$, without doing any fine-tuning
on the coupling constants, is to have really heavy masses for the scalar bosons
which mediate the $FCNC$. \cite{barenboim1, deshpande, barenboim3, fcnc2}
However this is not always possible and it depends on the values of the
parameters of the scalar potential as well as on the values that the
spontaneous $CP$ phases take.

The aim of this article is to study the relationship between the amount of
the $FCNC$ and the values taken for the spontaneous $CP$ phases $\alpha $
and $\theta $. To do this, we explore four cases in which we have values of $%
0$ or $\pi /2$ for the two $CP$ phases. Two of these cases (what we are
going to call $I$ and $IV$) were already studied in the literature, \cite
{barenboim1, deshpande} and the other two $\left( II\textbf{ and }III\right) $
are novel. As we shall see later, $\alpha $ is associated with the quark
sector and $\theta $ with the lepton sector. Therefore, we study four
combinations of maximal and no $CP$ violation in both sectors. What we find
is that, in order to suppress the $FCNC$ in the $LRSM$, we have to adjust  
\textit{only} the $CP$-violating phase of the quark sector. So, we may have
any value for the other $CP$-violating phase, implying the possibility to
obtain a large $CP$ violation in the lepton sector, which is of great
importance for current and future experiments. \cite{cplepton} In addition,
we find other scalar particles at the electroweak scale, different from the
Higgs boson of the $SM$, which are important in the phenomenology of the
model, and that, analogously to the case of $CP$ violation in the lepton
sector, are the focus of most current and future experiments. \cite{exp}

The outline of the paper is the following: In Section $2$ we present the
main features of the $LRSM$. In Section $3$ we show how the $FCNC$ arise in
the $LRSM$. In Section $4$ we present the four cases described above
together with the minimization conditions taken for the scalar potential. We
analyze their relevance and viability with respect to the $FCNC$. Finally in
Section $5$ we conclude. In Appendix $A$ we show the complete mass matrices
for the scalar particles.\ In Appendix $B$ we present how to get the minimum
possible order of magnitude for the masses of the flavour changing scalar
bosons, by means of experimental constraints from neutrinos.

\seCtion{The model}

The $LRSM$ is based on the group of symmetries $SU\left( 3\right) _C\otimes
SU(2)_L\otimes SU\left( 2\right) _R\otimes U(1)_{B-L}\otimes C\otimes P$,
where the discrete parity $\left( P\right) $ symmetry stands for the same
coupling constant $g$ for the $SU(2)_L$ and the $SU\left( 2\right) _R$
groups. Additionally, the discrete charge-parity $\left( CP\right) $
symmetry assures that there is no explicit $CP$ violation. Therefore we have
to search for it in a spontaneous way.

According to the left-right symmetry, quarks and leptons are placed in left-
and right-handed doublets: \cite{leftright, spontaneous}

\begin{eqnarray}
\Psi _{iL} &=&\left(  
\begin{array}{c}
\nu _i \\  
e_i
\end{array}
\right) _L\equiv \left( 2,1,-1\right) ,\hspace{3mm}\Psi _{iR}=\left(  
\begin{array}{c}
\nu _i \\  
e_i
\end{array}
\right) _R\equiv \left( 1,2,-1\right) ,  \nonumber \\
Q_{iL} &=&\left(  
\begin{array}{c}
u_i \\  
d_i
\end{array}
\right) _L\equiv \left( 2,1,\frac 13\right) ,\hspace{3mm}Q_{iR}=\left(  
\begin{array}{c}
u_i \\  
d_i
\end{array}
\right) _R\equiv \left( 1,2,\frac 13\right) ,
\end{eqnarray}
where $i=1,2,3$ is the generation index, and the representation with respect
to the gauge group is explicitly given.

The gauge bosons consist of two triplets:

\begin{equation}
\mathbf{W}_{\mu L}=\left(  
\begin{array}{c}
W_\mu ^{+} \\  
Z_\mu ^0 \\  
W_\mu ^{-}
\end{array}
\right) _L\equiv \left( 3,1,0\right) ,\hspace{3mm}\mathbf{W}_{\mu R}=\left(  
\begin{array}{c}
W_\mu ^{+} \\  
Z_\mu ^0 \\  
W_\mu ^{-}
\end{array}
\right) _R\equiv \left( 1,3,0\right) ,
\end{equation}
and one singlet:  
\begin{equation}
\mathbf{B}_\mu =B_\mu ^0\equiv \left( 1,1,0\right) .
\end{equation}

As both quarks and leptons are placed in doublets, we need a bidoublet of
scalar bosons to implement the symmetry breaking mechanism:

\begin{equation}
\Phi =\left(  
\begin{array}{cc}
\phi _1^0 & \phi _1^{+} \\  
\phi _2^{-} & \phi _2^0
\end{array}
\right) \equiv \left( 2,2,0\right) .
\end{equation}
However, this bidoublet leads to the same masses for the left-weak and the
right-weak bosons with equal electric charge. To avoid this problem, we have
to extend the Higgs sector by introducing two triplets:

\begin{equation}
\Delta _L=\left(  
\begin{array}{cc}
\frac{\delta _L^{+}}{\sqrt{2}} & \delta _L^{++} \\  
\delta _L^0 & \frac{-\delta _L^{+}}{\sqrt{2}}
\end{array}
\right) \equiv (3,1,2),\hspace{3mm}\Delta _R=\left(  
\begin{array}{cc}
\frac{\delta _R^{+}}{\sqrt{2}} & \delta _R^{++} \\  
\delta _R^0 & \frac{-\delta _R^{+}}{\sqrt{2}}
\end{array}
\right) \equiv (1,3,2),
\end{equation}
which have the additional interesting feature of implementing the see-saw
mechanism. This mechanism reproduces the observed light masses of the
left-handed neutrinos, whilst giving experimentally compatible heavy masses
to the right-handed ones. \cite{deshpande}-\cite{khasanov}

The symmetry breaking pattern of the bidoublet and the triplets is given by:

\begin{eqnarray}
\left\langle \Phi \right\rangle  &=&\frac 1{\sqrt{2}}\left(  
\begin{array}{cc}
k_1e^{i\alpha _1} & 0 \\  
0 & k_2e^{i\alpha _2}
\end{array}
\right) ,  \nonumber \\
\left\langle \Delta _L\right\rangle  &=&\frac 1{\sqrt{2}}\left(  
\begin{array}{cc}
0 & 0 \\  
\upsilon _Le^{i\theta _L} & 0
\end{array}
\right) ,  \nonumber \\
\left\langle \Delta _R\right\rangle  &=&\frac 1{\sqrt{2}}\left(  
\begin{array}{cc}
0 & 0 \\  
\upsilon _Re^{i\theta _R} & 0
\end{array}
\right) ,  \label{vevs}
\end{eqnarray}
where $k_1,k_2,\upsilon _L,\upsilon _R,\alpha _1,\alpha _2,\theta _L,$and $%
\theta _R$ are real numbers. There are some constraints on the values that
the vacuum expectation values $k_1,k_2,\upsilon _L,$ and $\upsilon _R$ may
take: $\upsilon _L$ must be much smaller than $k_1$ and $k_2\footnote{$%
k^2=k_1^2+k_2^2\simeq \left( 246\textbf{ }GeV\right) ^2.$}$ to keep the well
known experimental condition $M_{W_L}^2/M_{Z_L}^2\simeq \cos ^2\theta _W$.  
\cite{barenboim2, deshpande} In addition, $\upsilon _R$ must be at least $%
2.7\times 10^7$ $GeV$ to give really heavy masses to the right-weak bosons $%
W_R^{+},W_R^{-},$ and $Z_R^0$, and to fit the experimental constraints coming
from neutrinos.\footnote{%
See Appendix B.} \cite{barenboim2, deshpande, khasanov}

Under unitary transformations of the fermionic fields, the scalar fields
transform according to the relations:

\begin{eqnarray}
\Phi &\longrightarrow &U_L\Phi U_R^{\dagger },  \nonumber \\
\Delta _L &\longrightarrow &U_L\Delta _L U_L^{\dagger },  \nonumber \\
\Delta _R &\longrightarrow &U_R\Delta _R U_R^{\dagger },
\end{eqnarray}
where we can absorbe some of the phases of the scalar fields by defining:

\begin{eqnarray}
U_L &=&\left(  
\begin{array}{cc}
e^{i\gamma _L} & 0 \\  
0 & e^{-i\gamma _L}
\end{array}
\right) ,  \nonumber \\
U_R &=&\left(  
\begin{array}{cc}
e^{i\gamma _R} & 0 \\  
0 & e^{-i\gamma _R}
\end{array}
\right) ,  \nonumber \\
\gamma _L &=&\frac{\theta _L}2,  \nonumber \\
\gamma _R &=&\gamma _L-\alpha _2.
\end{eqnarray}
By means of these definitions, we are left with two genuine phases which we
call $\alpha $ and $\theta $:

\begin{eqnarray}
\left\langle \Phi \right\rangle  &=&\frac 1{\sqrt{2}}\left(  
\begin{array}{cc}
k_1e^{i\alpha } & 0 \\  
0 & k_2
\end{array}
\right) ,  \nonumber \\
\left\langle \Delta _L\right\rangle  &=&\frac 1{\sqrt{2}}\left(  
\begin{array}{cc}
0 & 0 \\  
\upsilon _L & 0
\end{array}
\right) ,  \nonumber \\
\left\langle \Delta _R\right\rangle  &=&\frac 1{\sqrt{2}}\left(  
\begin{array}{cc}
0 & 0 \\  
\upsilon _Re^{i\theta } & 0
\end{array}
\right) .  \label{gvevs}
\end{eqnarray}
The Lagrangian is $CP$ invariant by definition, so the only sources of $CP$
violation in the $LRSM$ are these two phases, obtained spontaneously.

The Yukawa terms in the Lagrangian are given by:

\begin{eqnarray}
-\frak{L}_Y^l &=&\sum_{a,b}h_{ab}^l\overline{\Psi _{aL}}\Phi \Psi _{bR}+%
\widetilde{h}_{ab}^l\overline{\Psi _{aL}}\widetilde{\Phi }\Psi _{bR}  
\nonumber  \label{leptonyukawa} \\
&&+if_{ab}\left[ \Psi _{aL}^TC\tau _2\Delta _L\Psi _{bL}+\left(
L\leftrightarrow R\right) \right] +h.c.,  \label{leptonyukawa}
\end{eqnarray}
for the leptons, and by:

\begin{equation}
-\frak{L}_Y^q=\sum_{a,b}h_{ab}^q\overline{Q_{aL}}\Phi Q_{bR}+\widetilde{h}%
_{ab}^q\overline{Q_{aL}}\widetilde{\Phi }Q_{bR}+h.c.,  \label{quarkyukawa}
\end{equation}
for the quarks, where $h^{l,q}$, $\widetilde{h}^{l,q}$, and $f$ are the
Yukawa coupling matrices, $\widetilde{\Phi }=\tau _2\Phi ^{*}\tau _2$, $C$
is the Dirac's charge-conjugation matrix, and $a,b$ label different
generations. Looking at the above expressions, we can see that the
spontaneous $CP$ phase $\theta $ enters only in the lepton sector whilst the
spontaneous $CP$ phase $\alpha $ enters in both the lepton and the quark
sector. Therefore, the phase in the $CKM$ matrix that is responsible for the
observed $CP$ violation in the quark sector, is function of $\alpha $ but
not of $\theta $. Then, once we have adjusted $\alpha $ to be consistent
with the experimental data, the amount of $CP$ violation in the lepton
sector will be determined only by $\theta $. These facts lead us to relate
the spontaneous $CP$ phase $\alpha $ with the quark sector, and the
spontaneous $CP$ phase $\theta $ with the lepton sector, although $\alpha $
will be responsible of a possible $CP$ violation in the lepton sector too.

The complete Lagrangian must be invariant under the transformations:

\begin{eqnarray}
\Psi _L &\leftrightarrow &\Psi _R,\hspace{3mm}Q_L\leftrightarrow Q_R,  
\nonumber \\
\Delta _L &\leftrightarrow &\Delta _R,\hspace{3mm}\Phi \leftrightarrow \Phi
^{\dagger },  \label{req}
\end{eqnarray}
due to the left-right symmetry requirements.

Thus, the most general scalar potential can be written as: \cite{barenboim1}-%
\cite{barenboim3}

\begin{equation}
\mathbf{V=V}_\Phi +\mathbf{V}_\Delta +\mathbf{V}_{\Phi \Delta },
\label{scalar potential}
\end{equation}
with

\begin{eqnarray*}
\mathbf{V}_\Phi &=&-\mu _1^2Tr\left( \Phi ^{\dagger }\Phi \right) -\mu
_2^2\left[ Tr\left( \widetilde{\Phi }\Phi ^{\dagger }\right) +Tr\left(  
\widetilde{\Phi }^{\dagger }\Phi \right) \right] \\
&&+\lambda _1\left[ Tr\left( \Phi \Phi ^{\dagger }\right) \right] ^2+\lambda
_2\left\{ \left[ Tr\left( \widetilde{\Phi }\Phi ^{\dagger }\right) \right]
^2+\left[ Tr\left( \widetilde{\Phi }^{\dagger }\Phi \right) \right]
^2\right\} \\
&&+\lambda _3\left[ Tr\left( \widetilde{\Phi }\Phi ^{\dagger }\right)
Tr\left( \widetilde{\Phi }^{\dagger }\Phi \right) \right] \\
&&+\lambda _4\left\{ Tr\left( \Phi ^{\dagger }\Phi \right) \left[ Tr\left(  
\widetilde{\Phi }\Phi ^{\dagger }\right) +Tr\left( \widetilde{\Phi }%
^{\dagger }\Phi \right) \right] \right\} ,
\end{eqnarray*}

\begin{eqnarray*}
\mathbf{V}_\Delta &=&-\mu _3^2\left[ Tr\left( \Delta _L\Delta _L^{\dagger
}\right) +Tr\left( \Delta _R\Delta _R^{\dagger }\right) \right] \\
&&+\rho _1\left\{ \left[ Tr\left( \Delta _L\Delta _L^{\dagger }\right)
\right] ^2+\left[ Tr\left( \Delta _R\Delta _R^{\dagger }\right) \right]
^2\right\} \\
&&+\rho _2\left[ Tr\left( \Delta _L\Delta _L\right) Tr\left( \Delta
_L^{\dagger }\Delta _L^{\dagger }\right) +Tr\left( \Delta _R\Delta _R\right)
Tr\left( \Delta _R^{\dagger }\Delta _R^{\dagger }\right) \right] \\
&&+\rho _3\left[ Tr\left( \Delta _L\Delta _L^{\dagger }\right) Tr\left(
\Delta _R\Delta _R^{\dagger }\right) \right] \\
&&+\rho _4\left[ Tr\left( \Delta _L\Delta _L\right) Tr\left( \Delta
_R^{\dagger }\Delta _R^{\dagger }\right) +Tr\left( \Delta _L^{\dagger
}\Delta _L^{\dagger }\right) Tr\left( \Delta _R\Delta _R\right) \right] ,
\end{eqnarray*}

\begin{eqnarray}
\mathbf{V}_{\Phi \Delta } &=&\alpha _1\left\{ Tr\left( \Phi ^{\dagger }\Phi
\right) \left[ Tr\left( \Delta _L\Delta _L^{\dagger }\right) +Tr\left(
\Delta _R\Delta _R^{\dagger }\right) \right] \right\}  \nonumber \\
&&+\alpha _2\{Tr\left( \widetilde{\Phi }^{\dagger }\Phi \right) Tr\left(
\Delta _R\Delta _R^{\dagger }\right) +Tr\left( \widetilde{\Phi }\Phi
^{\dagger }\right) Tr\left( \Delta _L\Delta _L^{\dagger }\right)  \nonumber
\\
&&+Tr\left( \widetilde{\Phi }\Phi ^{\dagger }\right) Tr\left( \Delta
_R\Delta _R^{\dagger }\right) +Tr\left( \widetilde{\Phi }^{\dagger }\Phi
\right) Tr\left( \Delta _L\Delta _L^{\dagger }\right) \}  \nonumber \\
&&+\alpha _3\left[ Tr\left( \Phi \Phi ^{\dagger }\Delta _L\Delta _L^{\dagger
}\right) +Tr\left( \Phi ^{\dagger }\Phi \Delta _R\Delta _R^{\dagger }\right)
\right]  \nonumber \\
&&+\beta _1\left[ Tr\left( \Phi \Delta _R\Phi ^{\dagger }\Delta _L^{\dagger
}\right) +Tr\left( \Phi ^{\dagger }\Delta _L\Phi \Delta _R^{\dagger }\right)
\right]  \nonumber \\
&&+\beta _2\left[ Tr\left( \widetilde{\Phi }\Delta _R\Phi ^{\dagger }\Delta
_L^{\dagger }\right) +Tr\left( \widetilde{\Phi }^{\dagger }\Delta _L\Phi
\Delta _R^{\dagger }\right) \right]  \nonumber \\
&&+\beta _3\left[ Tr\left( \Phi \Delta _R\widetilde{\Phi }^{\dagger }\Delta
_L^{\dagger }\right) +Tr\left( \Phi ^{\dagger }\Delta _L\widetilde{\Phi }%
\Delta _R^{\dagger }\right) \right] ,
\end{eqnarray}
where we have written out each term completely to display the full parity
symme\-try.\footnote{%
The parameters $\alpha _1$ and $\alpha _2$ in the scalar potential are
different to the phases of $\Phi $ in eq. (\ref{vevs}).} Note that, as a
consequence of the discrete left-right symmetry, all the terms in the
potential are self-conjugate. Therefore, all the parameters must be real,
avoiding any explicit source of $CP$ violation.

\seCtion{The $FCNC$}

We are going to concentrate on the Yukawa terms for the quark sector, eq. (%
\ref{quarkyukawa}):

\begin{equation}
-\frak{L}_Y^q=\sum_{a,b}h_{ab}^q\overline{Q_{aL}}\Phi Q_{bR}+\widetilde{h}%
_{ab}^q\overline{Q_{aL}}\widetilde{\Phi }Q_{bR}+h.c.
\end{equation}
In this Lagrangian $Q$ denotes the flavour eigenstates. Introducing the
vacuum expectation values, eq. (\ref{gvevs}), into the Yukawa terms, we
obtain the following mass matrices for the up and down quarks:

\begin{eqnarray}
M_{ab}^u &=&\frac 1{\sqrt{2}}\left( h_{ab}^qk_1e^{i\alpha }+\widetilde{h}%
_{ab}^qk_2\right) ,  \nonumber  \label{masses} \\
M_{ab}^d &=&\frac 1{\sqrt{2}}\left( h_{ab}^qk_2+\widetilde{h}%
_{ab}^qk_1e^{-i\alpha }\right) .  \label{masses}
\end{eqnarray}

To diagonalize these mass matrices, we have to rotate the flavour
eigenstates into the mass eigenstates, which we are going to call $Q^0$:

\begin{eqnarray}
Q_L^u &=&U_LQ_L^{0u},\hspace{5mm}Q_R^u=U_RQ_R^{0u},  \nonumber \\
Q_L^d &=&V_LQ_L^{0d},\hspace{5mm}Q_R^d=V_RQ_R^{0d}.
\end{eqnarray}
In this way, we can write the non diagonal mass matrices in terms of the
diagonal ones as:

\begin{eqnarray}
M^u &=&U_LM_{diag}^uU_R^{\dagger },  \nonumber \\
M^d &=&V_LM_{diag}^dV_R^{\dagger }.
\end{eqnarray}

For $k_1^2\neq k_2^2$ and $k_{\pm }^2\equiv $ $k_1^2\pm k_2^2$ we can invert
equations (\ref{masses}) to solve for $h^q$ and $\widetilde{h}^q$ in terms
of the diagonal matrices for the up and down quarks:

\begin{eqnarray}
h^q &=&\frac{\sqrt{2}}{k_{-}^2}\left( k_1e^{-i\alpha
}U_LM_{diag}^uU_R^{\dagger }-k_2V_LM_{diag}^dV_R^{\dagger }\right) ,  
\nonumber \\
\widetilde{h}^q &=&\frac{\sqrt{2}}{k_{-}^2}\left(
-k_2U_LM_{diag}^uU_R^{\dagger }+k_1e^{i\alpha }V_LM_{diag}^dV_R^{\dagger
}\right) .
\end{eqnarray}

Due to the left-right transformation, eq. (\ref{req}), $h^q$ and $\widetilde{%
h}^q$ must be hermitian. Then, we can define the $CKM$ matrices for the $%
LRSM $ as:

\begin{eqnarray}
K_L &=&U_L^{\dagger }V_L,  \nonumber \\
K_R &=&U_R^{\dagger }V_R,
\end{eqnarray}
which are related through the relation:

\begin{equation}
K=K_L=K_R^{*}.
\end{equation}

We can now write the general interaction term for the quark mass eigenstates
with the neutral $\phi -$type Higgs fields:

\begin{eqnarray}
&&\frac{\sqrt{2}}{k_{-}^2}\overline{u_L}^0\left[ M_{diag}^u\left(
k_1e^{-i\alpha }\phi _1^0-k_2\phi _2^{0*}\right) +K_LM_{diag}^dK_R^{\dagger
}\left( -k_2\phi _1^0+k_1e^{i\alpha }\phi _2^{0*}\right) \right] u_R^0  
\nonumber \\
&&+\frac{\sqrt{2}}{k_{-}^2}\overline{d_L}^0\left[ M_{diag}^d\left(
k_1e^{i\alpha }\phi _1^{0*}-k_2\phi _2^0\right) +K_L^{\dagger
}M_{diag}^uK_R\left( -k_2\phi _1^{0*}+k_1e^{-i\alpha }\phi _2^0\right)
\right] d_R^0.  \nonumber \\
&&
\end{eqnarray}

Defining two new orthogonal neutral fields: \cite{barenboim1, deshpande}

\begin{eqnarray}
\phi _{+}^0 &=&\frac 1{\left| k_{+}\right| }\left( -k_2\phi
_1^0+k_1e^{i\alpha }\phi _2^{0*}\right) ,  \nonumber \\
\phi _{-}^0 &=&\frac 1{\left| k_{+}\right| }\left( k_1e^{-i\alpha }\phi
_1^0+k_2\phi _2^{0*}\right) ,
\end{eqnarray}
with inverse transformations:

\begin{eqnarray}
\phi _1^0 &=&\frac 1{\left| k_{+}\right| }\left( -k_2\phi
_{+}^0+k_1e^{i\alpha }\phi _{-}^0\right) ,  \nonumber \\
\phi _2^0 &=&\frac 1{\left| k_{+}\right| }\left( k_1e^{i\alpha }\phi
_{+}^{0*}+k_2\phi _{-}^{0*}\right) ,
\end{eqnarray}
it is possible to write the general interaction term as:

\begin{eqnarray}
&&\frac{\sqrt{2}}{k_{-}^2}\overline{u_L}^0\left[ \phi _{-}^0\frac{k_{-}^2}{%
\left| k_{+}\right| }M_{diag}^u+\phi _{+}^0\left( -2\frac{k_1e^{-i\alpha }k_2%
}{\left| k_{+}\right| }M_{diag}^u+\left| k_{+}\right|
K_LM_{diag}^dK_R^{\dagger }\right) \right] u_R^0  \nonumber \\
&&+\frac{\sqrt{2}}{k_{-}^2}\overline{d_L}^0\left[ \phi _{-}^{0*}\frac{k_{-}^2%
}{\left| k_{+}\right| }M_{diag}^d+\phi _{+}^{0*}\left( -2\frac{k_1e^{i\alpha
}k_2}{\left| k_{+}\right| }M_{diag}^d+\left| k_{+}\right| K_L^{\dagger
}M_{diag}^uK_R\right) \right] d_R^0,  \nonumber \\
&&
\end{eqnarray}
where we can see clearly that there are $FCNC$ in the $LRSM$ associated with
the $\phi _{+}^0$ boson.

One possible way to avoid these $FCNC$, without performing any fine-tuning
on the coupling constants or the vacuum expectation values, is to give a
really heavy mass to the $\phi _{+}^0$ boson. In the next section, we are
going to investigate four cases in which we have different values for the
spontaneous $CP$ phases $\alpha $ and $\theta $. The idea is to search for a
model with a really heavy mass for $\phi _{+}^0$, for example, of the order
of $\upsilon _R$ $\left( 10^7\textbf{ }GeV\right) $. If we are able to find
it, we will have a model with $FCNC$ supressed enough to be consistent with
the experimental constraints, and with the additional feature of having a
spontaneous origin for the observed $CP$ violation.

\seCtion{Spontaneous $CP$ phases and $FCNC$}

In Section 2 we noticed that the spontaneous $CP$ phase $\alpha $ is related
directly with the amount of $CP$ violation in the quark sector, whilst $%
\theta $ is related to the lepton sector. In this section, we are going to
investigate the effects on the mass spectrum of the scalar sector, and
therefore, on the $FCNC$, of giving maximal $CP$ violation in both the quark
and the lepton sector $\left( \alpha =\pi /2,\hspace{2mm}\theta =\pi
/2\right) $, maximal $CP$ violation in the quark sector and no $CP$
violation in the lepton sector $\left( \alpha =\pi /2,\hspace{2mm}\theta
=0\right) $, no $CP$ violation in the quark sector and maximal $CP$
violation in the lepton sector $\left( \alpha =0,\hspace{2mm}\theta =\pi
/2\right) $, and no $CP$ violation in both the quark and the lepton sector $%
\left( \alpha =0,\hspace{2mm}\theta =0\right) $. The idea is to find what
restrictions on the spontaneous $CP$ phases are needed to obtain an
experimentally consistent $LRSM$. To do that we need the components of the
scalar mass matrices presented in Appendix A. The four cases defined above
will be called cases $I$, $II$, $III$, and $IV$, respectively.

\subsection{Case $I$: $\alpha =\pi /2$ and $\theta =\pi /2$}

In this case, we have maximal $CP$ violation in both the quark and the
lepton sector. The minimization conditions arising from the scalar
potential, eq. (\ref{scalar potential}), are the following:

\begin{eqnarray}
\beta _2 &=&\beta _3\frac{k_2^2}{k_1^2},  \nonumber \\
\rho _1 &=&\frac{\rho _3}2+\beta _1\frac{k_1k_2}{2\upsilon _L\upsilon _R},  
\nonumber \\
\lambda _2 &=&\frac{\lambda _3}2-\alpha _3\frac{\upsilon _L^2+\upsilon _R^2}{%
8\left( k_2^2-k_1^2\right) }+\beta _1\frac{\upsilon _L\upsilon _R}{8k_1k_2},
\nonumber \\
\mu _1^2 &=&-2\left( 2\lambda _2-\lambda _3\right) k_2^2+\frac{\alpha _1}%
2\left( \upsilon _L^2+\upsilon _R^2\right) +\lambda _1\left(
k_1^2+k_2^2\right) +\beta _1\frac{k_2}{2k_1}\upsilon _L\upsilon _R,  
\nonumber \\
\mu _2^2 &=&\frac{\lambda _4}2\left( k_1^2+k_2^2\right) +\frac{\alpha _2}%
2\left( \upsilon _L^2+\upsilon _R^2\right) +\beta _2\frac{k_1}{2k_2}\upsilon
_L\upsilon _R,  \nonumber \\
\mu _3^2 &=&\frac 12\left[ \alpha _1\left( k_1^2+k_2^2\right) +\alpha
_3k_2^2+2\rho _1\left( \upsilon _L^2+\upsilon _R^2\right) \right] .
\end{eqnarray}

Introducing these minimization conditions into the neutral scalar mass
matrix, eq. (\ref{neutral matrix}), and going to the basis $\{\phi
_{-}^r,\phi _{+}^r,\delta _R^r,\delta _L^r,\phi _{-}^i,\phi _{+}^i,\delta
_R^i,\delta _L^i\}$ through the general rotation matrix:

\begin{equation}
R=\frac 1{\left| k_{+}\right| }\left(  
\begin{array}{cccccccc}
k_1\cos \alpha & k_2 & 0 & 0 & k_1\sin \alpha & 0 & 0 & 0 \\  
-k_2 & k_1\cos \alpha & 0 & 0 & 0 & k_1\sin \alpha & 0 & 0 \\  
0 & 0 & \left| k_{+}\right| & 0 & 0 & 0 & 0 & 0 \\  
0 & 0 & 0 & \left| k_{+}\right| & 0 & 0 & 0 & 0 \\  
-k_1\sin \alpha & 0 & 0 & 0 & k_1\cos \alpha & -k_2 & 0 & 0 \\  
0 & k_1\sin \alpha & 0 & 0 & -k_2 & -k_1\cos \alpha & 0 & 0 \\  
0 & 0 & 0 & 0 & 0 & 0 & \left| k_{+}\right| & 0 \\  
0 & 0 & 0 & 0 & 0 & 0 & 0 & \left| k_{+}\right|
\end{array}
\right) ,
\end{equation}
we find the following neutral scalar mass matrix where we only have put the
leading terms represented by generic symbols, and used the fact that $%
\upsilon _L\upsilon _R\sim k^2$, which results from avoiding fine-tuning of
most parameters of the scalar potential:\footnote{%
There is no way to avoid fine-tuning the parameters of the scalar potential.
We have chosen to fine-tune the $\mu ^2$ elements which permits to have
fine-tunings on the fewest parameters of the scalar potential.}

\begin{equation}
M^2=\left(  
\begin{array}{cccccccc}
\alpha \upsilon _R^2 & \left( \lambda +\beta \right) k^2 & 0 & \beta
k\upsilon _R & \beta k^2 & \alpha \upsilon _R^2 & \alpha k\upsilon _R & 0 \\  
\left( \lambda +\beta \right) k^2 & \alpha \upsilon _R^2 & 0 & \beta
k\upsilon _R & \beta k^2 & \beta k^2 & \alpha k\upsilon _R & \beta k\upsilon
_R \\  
0 & 0 & 0 & 0 & 0 & 0 & 0 & \beta k^2 \\  
\beta k\upsilon _R & \beta k\upsilon _R & 0 & \beta \upsilon _R^2 & 0 &  
\beta k\upsilon _R & \left( \rho _3+\beta \right) k^2 & 0 \\  
\beta k^2 & \beta k^2 & 0 & 0 & \beta k^2 & \beta k^2 & 0 & \beta k\upsilon
_R \\  
\alpha \upsilon _R^2 & \beta k^2 & 0 & \beta k\upsilon _R & \beta k^2 &  
\alpha \upsilon _R^2 & \alpha k\upsilon _R & \beta k\upsilon _R \\  
\alpha k\upsilon _R & \alpha k\upsilon _R & 0 & \left( \rho _3+\beta \right)
k^2 & 0 & \alpha k\upsilon _R & \left( \rho _3+\beta \right) \upsilon _R^2 &  
0 \\  
0 & \beta k\upsilon _R & \beta k^2 & 0 & \beta k\upsilon _R & \beta
k\upsilon _R & 0 & \beta \upsilon _R^2
\end{array}
\right) .
\end{equation}

Defining two new orthogonal fields:

\begin{eqnarray}
\phi _{+}^{+} &=&\frac 1{\left| k_{+}\right| }\left( -k_2\phi _1^{+}+k_1\phi
_2^{+}\right) ,  \nonumber \\
\phi _{-}^{+} &=&\frac 1{\left| k_{+}\right| }\left( k_1\phi _1^{+}+k_2\phi
_2^{+}\right) ,
\end{eqnarray}
and their corresponding rotation matrix $R_{+}$:

\begin{equation}
R_{+}=\frac 1{\left| k_{+}\right| }\left(  
\begin{array}{cccc}
k_1 & k_2 & 0 & 0 \\  
-k_2 & k_1 & 0 & 0 \\  
0 & 0 & \left| k_{+}\right| & 0 \\  
0 & 0 & 0 & \left| k_{+}\right|
\end{array}
\right) ,
\end{equation}
we can obtain the singly and doubly charged mass matrices in the basis $%
\{\phi _{-}^{+},\phi _{+}^{+},\\\delta _R^{+},\delta _L^{+}\}$ and $\{\delta
_R^{++},\delta _L^{++}\}$ respectively:

\begin{equation}
M^{2+}=\left(  
\begin{array}{cccc}
\alpha \upsilon _R^2 & \left( 1-i\right) \alpha \upsilon _R^2 & \left(
1+i\right) \alpha k\upsilon _R & \left( 1-i\right) \beta k\upsilon _R \\  
\left( 1+i\right) \alpha \upsilon _R^2 & \alpha \upsilon _R^2 & \left(
1-i\right) \alpha k\upsilon _R & \left( 1-i\right) \beta k\upsilon _R \\  
\left( 1-i\right) \alpha k\upsilon _R & \left( 1+i\right) \alpha k\upsilon _R
& \alpha k^2 & \left( 1+i\right) \beta k^2 \\  
\left( 1+i\right) \beta k\upsilon _R & \left( 1+i\right) \beta k\upsilon _R
& \left( 1-i\right) \beta k^2 & \beta \upsilon _R^2
\end{array}
\right) ,
\end{equation}

\vspace{1cm}

\begin{equation}
M^{2++}=\left(  
\begin{array}{cc}
\rho \upsilon _R^2 & \left[ \left( 1+i\right) \beta +i\rho \right] k^2 \\  
\left[ \left( 1-i\right) \beta -i\rho \right] k^2 & \beta \upsilon _R^2
\end{array}
\right) .
\end{equation}

We performed a detailed numerical analysis where we chose an order of
magnitude for $\upsilon _R$ of $10^7$ $GeV$, according to the experimental
constraints on neutrinos,\footnote{%
See Appendix B.} and a value of $0.7$ for the free dimensionless parameters
of the potential. What we found is that, in this model with maximal $CP$
violation in both the quark and the lepton sector, the neutral scalar boson $%
\phi _F^0$ which contains a significant admixture of $\phi _{+}^i$ is
light enough\footnote{%
Of the order of $k=246$ $GeV$.} to allow for large $FCNC$. Therefore, this
kind of model is experimentaly unacceptable. Tables $1$, $2$, and $3$, show
the normalized components of the mass eigenstates and the corresponding
order of magnitude for the masses. Some previous conclusions about the state
of this case were presented in \cite{barenboim1}. Since there were a lot of
typos there, these conclusions were erroneous. One of the authors of \cite
{barenboim1} noted this mistake in \cite{barenboim3}.

\subsection{Case $II$: $\alpha =\pi /2$ and $\theta =0$}

In this case, we have maximal $CP$ violation in the quark sector and no $CP$
violation in the lepton sector. The minimization conditions arising from the
scalar potential, eq. (\ref{scalar potential}), are the following:

\begin{eqnarray}
\beta _2 &=&\frac 1{k_1^2}\left[ \beta _3k_2^2-\left( 2\rho _1-\rho
_3\right) \upsilon _L\upsilon _R\right] ,  \nonumber \\
\lambda _2 &=&\frac{\lambda _3}2-\frac 1{4\left( k_2^2-k_1^2\right) }\left[  
\frac{\alpha _3}2\left( \upsilon _L^2+\upsilon _R^2\right) +\left( \beta
_2+\beta _3\right) \upsilon _L\upsilon _R\right] ,  \nonumber \\
\mu _1^2 &=&-2\left( 2\lambda _2-\lambda _3\right) k_2^2+\frac{\alpha _1}%
2\left( \upsilon _L^2+\upsilon _R^2\right) +\lambda _1\left(
k_1^2+k_2^2\right) -\beta _2\upsilon _L\upsilon _R,  \nonumber \\
\mu _2^2 &=&\frac{\lambda _4}2\left( k_1^2+k_2^2\right) +\frac{\alpha _2}%
2\left( \upsilon _L^2+\upsilon _R^2\right) +\frac{\beta _1}4\upsilon
_L\upsilon _R,  \nonumber \\
\mu _3^2 &=&\frac 12\left[ \alpha _1\left( k_1^2+k_2^2\right) +\alpha
_3k_2^2+2\rho _1\left( \upsilon _L^2+\upsilon _R^2\right) \right] .
\end{eqnarray}

Introducing these minimization conditions into the neutral scalar mass
matrix, eq. (\ref{neutral matrix}), and going to the basis $\{\phi
_{-}^r,\phi _{+}^r,\delta _R^r,\delta _L^r,\phi _{-}^i,\phi _{+}^i,\delta
_R^i,\delta _L^i\}$ and $\{\phi _{-}^{+},\phi _{+}^{+},\delta _R^{+},\newline%
\delta _L^{+}\}$ through the rotation matrices $R$ and $R_{+}$ respectively,
we find the following neutral, singly charged, and doubly charged scalar
mass matrices:

\begin{equation}
M^2=\left(  
\begin{array}{cc}
M_{B11}^2 & M_{B12}^2 \\  
M_{B12}^{2\hspace{2mm}T} & M_{B22}^2
\end{array}
\right) ,
\end{equation}

\vspace{1cm}

\begin{equation}
M_{B11}^2=\left(  
\begin{array}{cccc}
\alpha \upsilon _R^2 & \left( \lambda +\beta \right) k^2 & \alpha k\upsilon
_R & \left( 2\rho _1-\rho _3\right) k\upsilon _R \\  
\left( \lambda +\beta \right) k^2 & \alpha \upsilon _R^2 & \alpha k\upsilon
_R & \beta k\upsilon _R \\  
\alpha k\upsilon _R & \alpha k\upsilon _R & 2\rho _1\upsilon _R^2 & \left(
2\rho _1+\rho _3\right) k^2/2 \\  
\left( 2\rho _1-\rho _3\right) k\upsilon _R & \beta k\upsilon _R & \left(
2\rho _1+\rho _3\right) k^2/2 & \left( \rho _3-2\rho _1\right) \upsilon
_R^2/2
\end{array}
\right) ,
\end{equation}

\vspace{1cm}

\begin{equation}
M_{B12}^2=\left(  
\begin{array}{cccc}
\beta k^2 & \alpha \upsilon _R^2 & 0 & \beta k\upsilon _R \\  
\left[ \beta +\rho _3-2\rho _1\right] k^2 & \beta k^2 & 0 & \left[ \beta
+\rho _3-2\rho _1\right] k\upsilon _R \\  
0 & \alpha k\upsilon _R & 0 & \beta k^2 \\  
\beta k\upsilon _R & \left[ \beta +\rho _3-2\rho _1\right] k\upsilon _R &  
\beta k^2 & 0
\end{array}
\right) ,
\end{equation}

\vspace{1cm}

\begin{equation}
M_{B22}^2=\left(  
\begin{array}{cccc}
2\left( \rho _3-2\rho _1\right) k^2 & \beta k^2 & 0 & \left( \rho _3-2\rho
_1\right) k\upsilon _R \\  
\beta k^2 & \alpha \upsilon _R^2 & 0 & \beta k\upsilon _R \\  
0 & 0 & 0 & \left( 2\rho _1-\rho _3\right) k^2/2 \\  
\left( \rho _3-2\rho _1\right) k\upsilon _R & \beta k\upsilon _R & \left(
2\rho _1-\rho _3\right) k^2/2 & \left( \rho _3-2\rho _1\right) \upsilon
_R^2/2
\end{array}
\right) ,
\end{equation}

\vspace{1cm}

\begin{equation}
M^{2+}=\left(  
\begin{array}{cc}
M_{B11}^{2+} & M_{B12}^{2+} \\  
M_{B12}^{2+\hspace{2mm}T} & M_{B22}^{2+}
\end{array}
\right) ,
\end{equation}

\vspace{1cm}

\begin{equation}
M_{B11}^{2+}=\left(  
\begin{array}{cc}
\alpha \upsilon _R^2 & -i\alpha \upsilon _R^2 \\  
i\alpha \upsilon _R^2 & \alpha \upsilon _R^2
\end{array}
\right) ,
\end{equation}

\vspace{1cm}

\begin{equation}
M_{B12}^{2+}=\left(  
\begin{array}{cc}
\left( 1-i\right) \alpha k\upsilon _R & i\left[ \left( \left( \rho _3-2\rho
_1\right) /\sqrt{2}\right) -\left( 1-i\right) \beta \right] k\upsilon _R \\  
\left( 1+i\right) \alpha k\upsilon _R & \left[ \left( 1+i\right) \beta
-i\left( \left( \rho _3-2\rho _1\right) /\sqrt{2}\right) \right] k\upsilon _R
\end{array}
\right) ,
\end{equation}

\vspace{1cm}

\begin{equation}
M_{B22}^{2+}=\left(  
\begin{array}{cc}
\alpha k^2 & \left[ i\left( \left( 2\rho _1-\rho _3\right) /2\right) +\left(
1-i\right) \beta \right] k^2 \\  
\left[ -i\left( \left( 2\rho _1-\rho _3\right) /2\right) +\left( 1-i\right)
\beta \right] k^2 & 2\left( \rho _3-2\rho _1\right) \upsilon _R^2
\end{array}
\right) ,
\end{equation}

\vspace{1cm}

\begin{equation}
M^{2++}=\left(  
\begin{array}{cc}
\rho \upsilon _R^2 & \left[ \left( 1+i\right) \beta +\rho +\left( \rho
_3-2\rho _1\right) /2\right] k^2 \\  
\left[ \left( 1-i\right) \beta +\rho +\left( \rho _3-2\rho _1\right)
/2\right] k^2 & \left( \rho _3-2\rho _1\right) \upsilon _R^2/2
\end{array}
\right) .
\end{equation}

In our numerical analysis we also found a neutral scalar boson $\phi _F^0$
containing a significant admixture of $\phi _{+}^i$, which tells us that
this model is also unacceptable. What we can conclude is that, independently
of the values that the spontaneous $CP$ phase $\theta $ takes, we will not
be able to obtain an experimentally consistent $LRSM$ if the amount of $CP$
violation in the quark sector is maximal. Tables $4$, $5$, and $6$, show the
normalized components of the mass eigenstates and the corresponding order of
magnitude for the masses.

\subsection{Case $III$: $\alpha =0$ and $\theta =\pi /2$}

In this case, we have no $CP$ violation in the quark sector and maximal $CP$
violation in the lepton sector. The minimization conditions arising from the
scalar potential, eq. (\ref{scalar potential}), are the following:

\begin{eqnarray}
\rho _1 &=&\frac{\rho _3}2,  \nonumber \\
\beta _2 &=&-\frac 1{k_1^2}\left( \beta _1k_1k_2+\beta _3k_2^2\right) ,  
\nonumber \\
\mu _1^2 &=&\lambda _1\left( k_1^2+k_2^2\right) +2\lambda _4k_1k_2+\frac{%
\alpha _1\left( k_1^2-k_2^2\right) -\alpha _3k_2^2}{2\left(
k_1^2-k_2^2\right) }\left( \upsilon _L^2+\upsilon _R^2\right) ,  \nonumber \\
\mu _2^2 &=&\left( 2\lambda _2+\lambda _3\right) k_1k_2+\frac{\lambda _4}%
2\left( k_1^2+k_2^2\right) +\frac{2\alpha _2\left( k_1^2-k_2^2\right)
+\alpha _3k_1k_2}{4\left( k_1^2-k_2^2\right) }\left( \upsilon _L^2+\upsilon
_R^2\right) ,  \nonumber \\
\mu _3^2 &=&\frac 12\left[ \alpha _1\left( k_1^2+k_2^2\right) +4\alpha
_2k_1k_2+\alpha _3k_2^2+2\rho _1\left( \upsilon _L^2+\upsilon _R^2\right)
\right] .
\end{eqnarray}

The neutral, singly charged, and doubly charged scalar mass matrices in the
new basis are:

\begin{equation}
M^2=\left(  
\begin{array}{cccccccc}
\lambda k^2 & \lambda k^2 & 0 & 0 & \beta k^2 & \beta k^2 & \alpha k\upsilon
_R & 0 \\  
\lambda k^2 & \alpha \upsilon _R^2 & 0 & 0 & \beta k^2 & \beta k^2 & \alpha
k\upsilon _R & \beta k\upsilon _R \\  
0 & 0 & 0 & 0 & 0 & 0 & 0 & 0 \\  
0 & 0 & 0 & 0 & 0 & \beta k\upsilon _R & \rho _3k^2 & 0 \\  
\beta k^2 & \beta k^2 & 0 & 0 & 0 & 0 & 0 & 0 \\  
\beta k^2 & \beta k^2 & 0 & \beta k\upsilon _R & 0 & \alpha \upsilon _R^2 & 0
& 0 \\  
\alpha k\upsilon _R & \alpha k\upsilon _R & 0 & \rho _3k^2 & 0 & 0 & \rho
_3\upsilon _R^2 & 0 \\  
0 & \beta k\upsilon _R & 0 & 0 & 0 & 0 & 0 & 0
\end{array}
\right) .
\end{equation}

\vspace{1cm}

\begin{equation}
M^{2+}=\left(  
\begin{array}{cccc}
\alpha \upsilon _R^2 & 0 & i\alpha k\upsilon _R & i\beta k\upsilon _R \\  
0 & 0 & 0 & 0 \\  
-i\alpha k\upsilon _R & 0 & \alpha k^2 & \beta k^2 \\  
-i\beta k\upsilon _R & 0 & \beta k^2 & \alpha k^2
\end{array}
\right) ,
\end{equation}

\vspace{1cm}

\begin{equation}
M^{2++}=\left(  
\begin{array}{cc}
\rho \upsilon _R^2 & \left( \beta +i\rho \right) k^2 \\  
\left( \beta -i\rho \right) k^2 & \alpha k^2
\end{array}
\right) .
\end{equation}

When we perfomed our numerical analysis in this kind of model, we found that
every neutral scalar eigenstate containing a significant admixture of $\phi
_{+}^0$, real or imaginary, has a mass of the order of $\upsilon _R$. Since
the minimum value that $\upsilon _R$ can take is $2.7\times 10^7$ $GeV$,
this is a model with $FCNC$ that are consistent with the experimental
constraints. It is interesting to note that, in addition to the field
analogous of the Higgs boson of the $SM$, two other neutral scalar particles%
\footnote{%
These non-$SM$ neutral scalar bosons contain a negligible admixture of $\phi
_{+}^0$, so their contributions to the $FCNC$ are suppressed.}, one singly
charged scalar particle, and one doubly charged scalar particle appear with
masses at the electroweak scale. There is no experimental constraint to this
order of magnitude for the masses of the new scalar particles. \cite{pdg,
newp} This is very interesting because the search for new scalar particles
is the focus of most current and future experiments. \cite{exp} Any
experimental evidence about this issue could give us some light about the
validity and viability of this class of models with left-right symmetries.
Tables $7$, $8$, and $9$, show the normalized components of the mass
eigenstates and the corresponding order of magnitude for the masses.

\subsection{Case $IV$: $\alpha =0$ and $\theta =0$}

In this case, we have no $CP$ violation in both the quark and the lepton
sector. The minimization conditions arising from the scalar potential, eq. (%
\ref{scalar potential}), are the following:

\begin{eqnarray}
\beta _2 &=&\frac 1{k_1^2}\left( -\beta _1k_1k_2-\beta _3k_2^2+\left( 2\rho
_1-\rho _3\right) \upsilon _L\upsilon _R\right) ,  \nonumber \\
\mu _1^2 &=&\lambda _1\left( k_1^2+k_2^2\right) +2\lambda _4k_1k_2  \nonumber
\\
&&+\frac 1{2\left( k_1^2-k_2^2\right) }\left[ 2(\beta _2k_1^2-\beta
_3k_2^2)\upsilon _L\upsilon _R+\left[ \alpha _1\left( k_1^2-k_2^2\right)
-\alpha _3k_2^2\right] \left( \upsilon _L^2+\upsilon _R^2\right) \right] ,  
\nonumber \\
\mu _2^2 &=&\left( 2\lambda _2+\lambda _3\right) k_1k_2+\frac{\lambda _4}%
2\left( k_1^2+k_2^2\right)  \nonumber \\
&&+\frac 1{4\left( k_1^2-k_2^2\right) }((\beta _1(k_1^2-k_2^2)-2k_1k_2(\beta
_2-\beta _3))\upsilon _L\upsilon _R  \nonumber \\
&&+\left[ 2\alpha _2\left( k_1^2-k_2^2\right) +\alpha _3k_1k_2\right] \left(
\upsilon _L^2+\upsilon _R^2\right) ),  \nonumber \\
\mu _3^2 &=&\frac 12\left[ \alpha _1\left( k_1^2+k_2^2\right) +4\alpha
_2k_1k_2+\alpha _3k_2^2+2\rho _1\left( \upsilon _L^2+\upsilon _R^2\right)
\right] .
\end{eqnarray}

The neutral, singly charged, and doubly charged scalar mass matrices in the
new basis are:

\begin{equation}
M^2=\left(  
\begin{array}{cc}
M_{B11}^2 & 0 \\  
0 & M_{B22}^2
\end{array}
\right) ,
\end{equation}

\vspace{1cm}

\begin{equation}
M_{B11}^2=\left(  
\begin{array}{cccc}
\lambda k^2 & \lambda k^2 & \alpha k\upsilon _R & \left( 2\rho _1-\rho
_3\right) k\upsilon _R \\  
\lambda k^2 & \alpha \upsilon _R^2 & \alpha k\upsilon _R & \beta k\upsilon _R
\\  
\alpha k\upsilon _R & \alpha k\upsilon _R & 2\rho _1\upsilon _R^2 & \left(
2\rho _1+\rho _3\right) k^2/2 \\  
\left( 2\rho _1-\rho _3\right) k\upsilon _R & \beta k\upsilon _R & \left(
2\rho _1+\rho _3\right) k^2/2 & \left( \rho _3-2\rho _1\right) \upsilon
_R^2/2
\end{array}
\right) ,
\end{equation}

\vspace{1cm}

\newpage
\[
M_{B22}^2=  
\]

\vspace{-5mm}

\begin{equation}
\left(  
\begin{array}{cccc}
2\left( \rho _3-2\rho _1\right) k^2 & \left[ \beta -2\left( \rho _3-2\rho
_1\right) \right] k^2 & 0 & \left( \rho _3-2\rho _1\right) k\upsilon _R \\  
\left[ \beta -2\left( \rho _3-2\rho _1\right) \right] k^2 & \alpha \upsilon
_R^2 & 0 & \left[ \beta -2\left( \rho _3-2\rho _1\right) \right] k\upsilon _R
\\  
0 & 0 & 0 & \left( 2\rho _1-\rho _3\right) k^2/2 \\  
\left( \rho _3-2\rho _1\right) k\upsilon _R & \left[ \beta -2\left( \rho
_3-2\rho _1\right) \right] k\upsilon _R & \left( 2\rho _1-\rho _3\right)
k^2/2 & \left( \rho _3-2\rho _1\right) \upsilon _R^2/2
\end{array}
\right) ,
\end{equation}

\vspace{1cm}

\begin{equation}
M^{2+}=\left(  
\begin{array}{cccc}
\alpha \upsilon _R^2 & \beta k^2 & \alpha k\upsilon _R & \beta k\upsilon _R
\\  
\beta k^2 & \left( \rho _3-2\rho _1\right) k^2 & 0 & \left[ \left( \rho
_3-2\rho _1\right) /\sqrt{2}\right] k\upsilon _R \\  
\alpha k\upsilon _R & 0 & \alpha k^2 & \left[ \beta +\left( \rho _3-2\rho
_1\right) /4\right] k^2 \\  
\beta k\upsilon _R & \left[ \left( \rho _3-2\rho _1\right) /\sqrt{2}\right]
k\upsilon _R & \left[ \beta +\left( \rho _3-2\rho _1\right) /4\right] k^2 &  
\left[ \left( \rho _3-2\rho _1\right) /2\right] \upsilon _R^2
\end{array}
\right) ,
\end{equation}

\vspace{1cm}

\begin{equation}
M^{2++}=\left(  
\begin{array}{cc}
\rho \upsilon _R^2 & \left( \beta +\rho -\rho _3+2\rho _1\right) k^2 \\  
\left( \beta +\rho -\rho _3+2\rho _1\right) k^2 & \left[ \left( \rho
_3-2\rho _1\right) /2\right] \upsilon _R^2
\end{array}
\right) .
\end{equation}

Finally, in this last case our numerical analysis leads us to the same
conclusions already obtained by Deshpande et. al.. \cite{deshpande} In this
model we find that all the non-$SM$ Higgs bosons have a mass of the order of $%
\upsilon _R$, avoiding any large $FCNC$. This model is exactly equal to the $%
SM$ in the limit in which $\upsilon _R$ goes to the infinity. If we compare
the scalar mass spectrum of the cases $III$ and $IV$, we find that the value
which $\theta $ takes only affects the order of the masses of the scalar
particles and not the amount of the $FCNC$. Thus, we can conclude from the
two last cases that, independently of the value that the spontaneous $CP$
phase $\theta $ takes, we will have an experimentally consistent $LRSM$ if
there is no $CP$ violation in the quark sector. To avoid an explicit origin
for the $CP$ violation in the quark sector, we have to adjust $\alpha $ to
be small enough so as not to change the main features and results found, and
to lead to the correct experimental value for the $CKM$ phase of the $SM$.
Effectively, to obtain the correct value for the $CKM$ phase of the $SM$, we
need a value close to zero for the spontaneous $CP$ phase $\alpha $. \cite
{spontaneous2} Tables $10$, $11$, and $12$, show the normalized components
of the mass eigenstates and the corresponding order of magnitude for the
masses.

\seCtion{Conclusions}

In this paper, we have presented a detailed analysis of the relationship
between the $FCNC$ and the two spontaneous $CP$ phases present in the $LRSM$
with one doublet and two triplets of scalar bosons. Such quantity of scalar
bosons is necessary in order to give phenomenologically acceptable masses
to the right-weak bosons, and to implement the see-saw mechanism. Each
spontaneous $CP$ phase is related with one sector of matter: the quark or
the leptonic one. Different combinations of maximal and null values between
the two spontaneous $CP$ phases lead us to four different cases
corresponding to maximal and/or no $CP$ violation in the quark and the
lepton sector. What we found is that, each one of these cases leads to a
different mass spectrum for the scalar bosons, allowing or avoiding $FCNC$.

The main result of this paper is that the only way to suppress the $FCNC$ is
to adjust close to zero the spontaneous $CP$ phase associated with the quark
sector, as in the cases $III$ and $IV$. Then, the amount of $CP$ violation
in the lepton sector does not depend on the theoretical restrictions studied
in this paper and, therefore, a possible large $CKM$-like phase for the
lepton sector will only be restricted by elementary-process constraints.
Additionally, in the case $III$, we have the possibility to observe new
scalar bosons at the electroweak scale, which is a source of theoretical
inspiration and many experimental works. The other non-$SM$ particles, and
their corresponding phenomenology, are at the scale of $10^7$ $GeV$, which
is well beyond the reach of the next generation of accelerators. This result
comes from restrictions arising from neutrino masses. From the cases $I$ and  
$II$ we can see that the $FCNC$ are present if the $CP$ violation in the
quark sector is maximal. Therefore, this kind of models are experimentally
unacceptable. We notice that our results were obtained in a general
framework, which has let us study the four cases presented above. The
results obtained in the case $IV$ agree with those previously found in \cite
{deshpande}, which implies the consistence of our results for the other
three cases.

The $LRSM$ is then viable; it gives a natural explanation to the origin of
the parity $\left( P\right) $ violation and the charge-parity $\left(
CP\right) $ violation too. It does not have large $FCNC$ that enter in
conflict with the experimental data, and gives us a rich phenomenological world
of $CP$ violation in the lepton sector and new Higgs particles, neutral,
singly charged, and doubly charged, at the electroweak scale.

\section*{Acknowledgements}

We want to acknowledge the Grupo de Campos y Part\'{\i}culas - Universidad
Nacional de Colombia for useful discussions, and Marta Losada and Jaime
Gaviria for comments on the manuscript. Y. R. wants to thank the Centro de
Investigaciones of the Universidad Antonio Nari\~{n}o for its support and
confidence. This work was supported by Fundaci\'{o}n Mazda para el Arte y la
Ciencia, and Universidad Nacional de Colombia through contract DIB 803629.

\appendix  

\seCtion{Mass matrices for the scalar particles}

In this appendix we show the mass matrices for the neutral, singly charged,
and doubly charged scalar particles. We present the results before
substituting the minimization conditions. Although these mass matrices have
already been given in \cite{barenboim1}, there are lots of typos there,
which led to erroneous conclusions. One of the authors of \cite{barenboim1}
noted this mistake in \cite{barenboim3}.

\subsection{Neutral scalar mass matrix}

The components of the symmetric neutral scalar mass matrix in the $\{\phi
_1^r,\phi _2^r,\delta _R^r,\delta _L^r,\\\phi _1^i,\phi _2^i,\delta
_R^i,\delta _L^i\}$ basis are:

\vspace{5mm}

$\emph{M}_{11}^2=-\mu _1^2+\lambda _1\left[ k_1^2\left( 2\cos ^2\alpha
+1\right) +k_2^2\right] +2\left( 2\lambda _2+\lambda _3\right)
k_2^2+6\lambda _4k_1k_2\cos \alpha ,$

\hspace{1.4cm}$+\frac 12\alpha _1\left( \upsilon _L^2+\upsilon _R^2\right)
+\beta _2\upsilon _L\upsilon _R\cos \theta ,$\vspace{3mm}

$\emph{M}_{12}^2=-2\mu _2^2+2\left( \lambda _1+4\lambda _2+2\lambda
_3\right) k_1k_2\cos \alpha +\lambda _4\left[ k_1^2\left( 2\cos ^2\alpha
+1\right) +3k_2^2\right] $

\hspace{1.4cm}$+\alpha _2\left( \upsilon _L^2+\upsilon _R^2\right) +\frac
12\beta _1\upsilon _L\upsilon _R\cos \theta ,$\vspace{3mm}

$\emph{M}_{13}^2=\alpha _1k_1\upsilon _R\cos \alpha \cos \theta +2\alpha
_2k_2\upsilon _R\cos \theta +\frac 12\upsilon _L\left( \beta _1k_2+2\beta
_2k_1\cos \alpha \right) ,$\vspace{3mm}

$\emph{M}_{14}^2=\alpha _1k_1\upsilon _L\cos \alpha +2\alpha _2k_2\upsilon
_L+\frac 12\upsilon _R\left( \beta _1k_2\cos \theta +2\beta _2k_1\cos \left(
\theta -\alpha \right) \right) ,$\vspace{3mm}

$\emph{M}_{15}^2=\lambda _1k_1^2\sin 2\alpha +2\lambda _4k_1k_2\sin \alpha
+\beta _2\upsilon _L\upsilon _R\sin \theta ,$\vspace{3mm}

$\emph{M}_{16}^2=-8\lambda _2k_1k_2\sin \alpha -\lambda _4k_1^2\sin 2\alpha
-\frac 12\beta _1\upsilon _L\upsilon _R\sin \theta ,$\vspace{3mm}

$\emph{M}_{17}^2=\alpha _1k_1\upsilon _R\cos \alpha \sin \theta +2\alpha
_2k_2\upsilon _R\sin \theta +\beta _2k_1\upsilon _L\sin \alpha ,$\vspace{3mm}

$\emph{M}_{18}^2=\frac 12\upsilon _R\left( \beta _1k_2\sin \theta +2\beta
_2k_1\sin \left( \theta -\alpha \right) \right) ,$\vspace{3mm}

$\emph{M}_{22}^2=-\mu _1^2+\lambda _1\left( k_1^2+3k_2^2\right) +2\left(
2\lambda _2\cos 2\alpha +\lambda _3\right) k_1^2+6\lambda _4k_1k_2\cos
\alpha $

\hspace{1.4cm}$+\frac 12\left( \alpha _1+\alpha _3\right) \left( \upsilon
_L^2+\upsilon _R^2\right) +\beta _3\upsilon _L\upsilon _R\cos \theta ,$%
\vspace{3mm}

$\emph{M}_{23}^2=\alpha _1k_2\upsilon _R\cos \theta +2\alpha _2k_1\upsilon
_R\cos \alpha \cos \theta +\alpha _3k_2\upsilon _R\cos \theta $

\hspace{1.4cm}$+\frac 12\upsilon _L\left( \beta _1k_1\cos \alpha +2\beta
_3k_2\right) ,$\vspace{3mm}

$\emph{M}_{24}^2=\left( \alpha _1+\alpha _3\right) k_2\upsilon _L+2\alpha
_2k_1\upsilon _L\cos \alpha +\frac 12\upsilon _R\left( \beta _1k_1\cos
\left( \theta -\alpha \right) +2\beta _3k_2\cos \theta \right) ,$\vspace{3mm}

$\emph{M}_{25}^2=2\left( \lambda _1-4\lambda _2+2\lambda _3\right)
k_1k_2\sin \alpha +\lambda _4k_1^2\sin 2\alpha +\frac 12\beta _1\upsilon
_L\upsilon _R\sin \theta ,$\vspace{3mm}

$\emph{M}_{26}^2=-4\lambda _2k_1^2\sin 2\alpha -2\lambda _4k_1k_2\sin \alpha
-\beta _2\upsilon _L\upsilon _R\sin \theta ,$\vspace{3mm}

$\emph{M}_{27}^2=\alpha _1k_2\upsilon _R\sin \theta +2\alpha _2k_1\upsilon
_R\cos \alpha \sin \theta +\alpha _3k_2\upsilon _R\sin \theta +\frac 12\beta
_1k_1\upsilon _L\sin \alpha ,$\vspace{3mm}

$\emph{M}_{28}^2=\frac 12\upsilon _R\left( \beta _1k_1\sin \left( \theta
-\alpha \right) +2\beta _3k_2\sin \theta \right) ,$\vspace{3mm}

$\emph{M}_{33}^2=-\mu _3^2+\frac 12\alpha _1\left( k_1^2+k_2^2\right)
+2\alpha _2k_1k_2\cos \alpha +\frac 12\alpha _3k_2^2+\rho _1\upsilon
_R^2\left( 2\cos ^2\theta +1\right) $

\hspace{1.4cm}$+\frac 12\rho _3\upsilon _L^2,$\vspace{3mm}

$\emph{M}_{34}^2=\frac 12\beta _1k_1k_2\cos \alpha +\frac 12\beta
_2k_1^2\cos 2\alpha +\frac 12\beta _3k_2^2+\rho _3\upsilon _L\upsilon _R\cos
\theta ,$\vspace{3mm}

$\emph{M}_{35}^2=\alpha _1k_1\upsilon _R\sin \alpha \cos \theta -\beta
_2k_1\upsilon _L\sin \alpha ,$\vspace{3mm}

$\emph{M}_{36}^2=-2\alpha _2k_1\upsilon _R\sin \alpha \cos \theta +\frac
12\beta _1k_1\upsilon _L\sin \alpha ,$\vspace{3mm}

$\emph{M}_{37}^2=\rho _1\upsilon _R^2\sin 2\theta ,$\vspace{3mm}

$\emph{M}_{38}^2=-\frac 12\beta _1k_1k_2\sin \alpha -\frac 12\beta
_2k_1^2\sin 2\alpha ,$\vspace{3mm}

$\emph{M}_{44}^2=-\mu _3^2+\frac 12\alpha _1\left( k_1^2+k_2^2\right)
+2\alpha _2k_1k_2\cos \alpha +\frac 12\alpha _3k_2^2+3\rho _1\upsilon
_L^2+\frac 12\rho _3\upsilon _R^2,$\vspace{3mm}

$\emph{M}_{45}^2=\alpha _1k_1\upsilon _L\sin \alpha +\frac 12\beta
_1k_2\upsilon _R\sin \theta +\beta _2k_1\upsilon _R\sin \left( \theta
-\alpha \right) ,$\vspace{3mm}

$\emph{M}_{46}^2=-2\alpha _2k_1\upsilon _L\sin \alpha -\frac 12\beta
_1k_1\upsilon _R\sin \left( \theta -\alpha \right) -\beta _3k_2\upsilon
_R\sin \theta ,$\vspace{3mm}

$\emph{M}_{47}^2=\frac 12\beta _1k_1k_2\sin \alpha +\frac 12\beta
_2k_1^2\sin 2\alpha +\rho _3\upsilon _L\upsilon _R\sin \theta ,$\vspace{3mm}

$\emph{M}_{48}^2=0,$\vspace{3mm}

$\emph{M}_{55}^2=-\mu _1^2+\lambda _1\left[ k_1^2\left( 2\sin ^2\alpha
+1\right) +k_2^2\right] +2\left( -2\lambda _2+\lambda _3\right)
k_2^2+2\lambda _4k_1k_2\cos \alpha $

\hspace{1.4cm}$+\frac 12\alpha _1\left( \upsilon _L^2+\upsilon _R^2\right)
-\beta _2\upsilon _L\upsilon _R\cos \theta ,$\vspace{3mm}

$\emph{M}_{56}^2=2\mu _2^2-8\lambda _2k_1k_2\cos \alpha -\lambda _4\left[
k_1^2\left( 2\sin ^2\alpha +1\right) +k_2^2\right] -\alpha _2\left( \upsilon
_L^2+\upsilon _R^2\right) $

\hspace{1.4cm}$+\frac 12\beta _1\upsilon _L\upsilon _R\cos \theta ,$%
\vspace{3mm}

$\emph{M}_{57}^2=\alpha _1k_1\upsilon _R\sin \alpha \sin \theta +\frac
12\beta _1k_2\upsilon _L+\beta _2k_1\upsilon _L\cos \alpha ,$\vspace{3mm}

$\emph{M}_{58}^2=-\frac 12\beta _1k_2\upsilon _R\cos \theta -\beta
_2k_1\upsilon _R\cos \left( \theta -\alpha \right) ,$\vspace{3mm}

$\emph{M}_{66}^2=-\mu _1^2+\lambda _1\left( k_1^2+k_2^2\right) +2\left(
-2\lambda _2\cos 2\alpha +\lambda _3\right) k_1^2+2\lambda _4k_1k_2\cos
\alpha $

\hspace{1.4cm}$+\frac 12\left( \alpha _1+\alpha _3\right) \left( \upsilon
_L^2+\upsilon _R^2\right) -\beta _3\upsilon _L\upsilon _R\cos \theta ,$%
\vspace{3mm}

$\emph{M}_{67}^2=-2\alpha _2k_1\upsilon _R\sin \alpha \sin \theta -\frac
12\beta _1k_1\upsilon _L\cos \alpha -\beta _3k_2\upsilon _L,$\vspace{3mm}

$\emph{M}_{68}^2=\frac 12\beta _1k_1\upsilon _R\cos \left( \theta -\alpha
\right) +\beta _3k_2\upsilon _R\cos \theta ,$\vspace{3mm}

$\emph{M}_{77}^2=-\mu _3^2+\frac 12\alpha _1\left( k_1^2+k_2^2\right)
+2\alpha _2k_1k_2\cos \alpha +\frac 12\alpha _3k_2^2+\rho _1\upsilon
_R^2\left( 2\sin ^2\theta +1\right) $

\hspace{1.4cm}$+\frac 12\rho _3\upsilon _L^2,$\vspace{3mm}

$\emph{M}_{78}^2=\frac 12\beta _1k_1k_2\cos \alpha +\frac 12\beta
_2k_1^2\cos 2\alpha +\frac 12\beta _3k_2^2,$\vspace{3mm}

$\emph{M}_{88}^2=-\mu _3^2+\frac 12\alpha _1\left( k_1^2+k_2^2\right)
+2\alpha _2k_1k_2\cos \alpha +\frac 12\alpha _3k_2^2+\rho _1\upsilon
_L^2+\frac 12\rho _3\upsilon _R^2.$

\begin{equation}  \label{neutral matrix}
\end{equation}

\subsection{Singly charged scalar mass matrix}

The components of the hermitian singly charged scalar mass matrix in the $%
\{\phi _1^{+},\phi _2^{+},\\\delta _R^{+},\delta _L^{+}\}$ basis are:

\vspace{5mm}

$\emph{M}_{11}^{+2}=-\mu _1^2+\lambda _1\left( k_1^2+k_2^2\right) +2\lambda
_4k_1k_2\cos \alpha +\frac 12\alpha _1\left( \upsilon _L^2+\upsilon
_R^2\right) +\frac 12\alpha _3\upsilon _R^2,$\vspace{3mm}

$\emph{M}_{12}^{+2}=2\mu _2^2-2\left( 2\lambda _2e^{i\alpha }+\lambda
_3e^{-i\alpha }\right) k_1k_2-\lambda _4\left( k_1^2+k_2^2\right) -\alpha
_2\left( \upsilon _L^2+\upsilon _R^2\right) ,$\vspace{3mm}

$\emph{M}_{13}^{+2}=\frac 1{2\sqrt{2}}\alpha _3k_1\upsilon _Re^{i\left(
\theta -\alpha \right) }-\frac 1{2\sqrt{2}}\upsilon _L\left( \beta
_1k_2+2\beta _2k_1e^{i\alpha }\right) ,$\vspace{3mm}

$\emph{M}_{14}^{+2}=\frac 1{2\sqrt{2}}\alpha _3k_2\upsilon _L+\frac 1{2\sqrt{%
2}}\upsilon _R\left( \beta _1k_1e^{i\left( \theta -\alpha \right) }+2\beta
_3k_2e^{i\theta }\right) ,$\vspace{3mm}

$\emph{M}_{22}^{+2}=-\mu _1^2+\lambda _1\left( k_1^2+k_2^2\right) +2\lambda
_4k_1k_2\cos \alpha +\frac 12\alpha _1\left( \upsilon _L^2+\upsilon
_R^2\right) +\frac 12\alpha _3\upsilon _L^2,$\vspace{3mm}

$\emph{M}_{23}^{+2}=\frac 1{2\sqrt{2}}\alpha _3k_2\upsilon _Re^{i\theta
}+\frac 1{2\sqrt{2}}\upsilon _L\left( \beta _1k_1e^{i\alpha }+2\beta
_3k_2\right) ,$\vspace{3mm}

$\emph{M}_{24}^{+2}=\frac 1{2\sqrt{2}}\alpha _3k_1\upsilon _Le^{i\alpha
}-\frac 1{2\sqrt{2}}\upsilon _R\left( \beta _1k_2e^{i\theta }+2\beta
_2k_1e^{i\left( \theta -\alpha \right) }\right) ,$\vspace{3mm}

$\emph{M}_{33}^{+2}=-\mu _3^2+\frac 12\alpha _1\left( k_1^2+k_2^2\right)
+2\alpha _2k_1k_2\cos \alpha +\frac 14\alpha _3\left( k_1^2+k_2^2\right)
+\rho _1\upsilon _R^2+\frac 12\rho _3\upsilon _L^2,$\vspace{3mm}

$\emph{M}_{34}^{+2}=\frac 14\beta _1\left( k_1^2+k_2^2\right) +\frac
12\left( \beta _2e^{-i\alpha }+\beta _3e^{i\alpha }\right) k_1k_2,$%
\vspace{3mm}

$\emph{M}_{44}^{+2}=-\mu _3^2+\frac 12\alpha _1\left( k_1^2+k_2^2\right)
+2\alpha _2k_1k_2\cos \alpha +\frac 14\alpha _3\left( k_1^2+k_2^2\right)
+\rho _1\upsilon _L^2+\frac 12\rho _3\upsilon _R^2.$

\begin{equation}  \label{singly matrix}
\end{equation}
\newpage

\subsection{Doubly charged scalar mass matrix}

The components of the hermitian doubly charged scalar mass matrix in the $%
\{\delta _R^{++},\\\delta _L^{++}\}$ basis are:

\vspace{5mm}

$\emph{M}_{11}^{++2}=-\mu _3^2+\frac 12\alpha _1\left( k_1^2+k_2^2\right)
+2\alpha _2k_1k_2\cos \alpha +\frac 12\alpha _3k_1^2+\left( \rho _1+2\rho
_2\right) \upsilon _R^2+\frac 12\rho _3\upsilon _L^2,$\vspace{3mm}

$\emph{M}_{12}^{++2}=2\rho _4\upsilon _L\upsilon _Re^{i\theta }+\frac
12\beta _1k_1k_2e^{i\alpha }+\frac 12\beta _2k_2^2+\frac 12\beta
_3k_1^2e^{2i\alpha },$\vspace{3mm}

$\emph{M}_{22}^{++2}=-\mu _3^2+\frac 12\alpha _1\left( k_1^2+k_2^2\right)
+2\alpha _2k_1k_2\cos \alpha +\frac 12\alpha _3k_1^2+\left( \rho _1+2\rho
_2\right) \upsilon _L^2+\frac 12\rho _3\upsilon _R^2.$

\begin{equation}  \label{double matrix}
\end{equation}

\seCtion{$\upsilon _R$ from neutrino physics}

In this appendix we are going to show how to obtain an order of magnitude
for $\upsilon _R$ from the experimental restrictions on neutrino masses.

The Yukawa terms in the Lagrangian for the lepton sector are given by:

\begin{eqnarray}
-\frak{L}_Y^l &=&\sum_{a,b}h_{ab}^l\overline{\Psi _{aL}}\Phi \Psi _{bR}+%
\widetilde{h}_{ab}^l\overline{\Psi _{aL}}\widetilde{\Phi }\Psi _{bR}  
\nonumber  \label{leptonyukawa} \\
&&+if_{ab}\left[ \Psi _{aL}^TC\tau _2\Delta _L\Psi _{bL}+\left(
L\leftrightarrow R\right) \right] +h.c.,  \label{yuklep}
\end{eqnarray}
where, as in the quark Yukawa case, $h^l$ and $\widetilde{h}^l$ must be
hermitian. For convenience, we will work with a single generation, and
ignore the spontaneous $CP$ phases. Introducing the vacuum expectation
values into eq. (\ref{yuklep}), we obtain the following mass terms:

\begin{equation}
\frac 1{\sqrt{2}}\left[ \left( h^lk_1+\widetilde{h}^lk_2\right) \overline{%
\nu _L}\nu _R+\left( h^lk_2+\widetilde{h}^lk_1\right) \overline{e_L}%
e_R+f\left( \upsilon _R\overline{\nu _R^c}\nu _R+\upsilon _L\overline{\nu
_L^c}\nu _L\right) \right] +h.c..
\end{equation}

Neutrino mass terms derive both from the $h^l$ and $\widetilde{h}^l$ terms,
which lead to a Dirac mass, and from the $f$ term, which leads to a Majorana
mass. Defining, as usual, $\psi ^c\equiv C\left( \overline{\psi }\right) ^T$%
, it is convenient to employ the self-conjugate spinors:

\begin{equation}
\nu =\frac 1{\sqrt{2}}\left( \nu _L+\nu _L^c\right) ,\hspace{5mm}N=\frac 1{%
\sqrt{2}}\left( \nu _R+\nu _R^c\right) .
\end{equation}
Thus, the neutrino mass terms can be written as:

\begin{equation}
\left(  
\begin{array}{cc}
\overline{\nu } & \overline{N}
\end{array}
\right) \left(  
\begin{array}{cc}
\sqrt{2}f\upsilon _L & h_Dk_{+} \\  
h_Dk_{+} & \sqrt{2}f\upsilon _R
\end{array}
\right) \left(  
\begin{array}{c}
\nu \\  
N
\end{array}
\right) ,  \label{neutrinomass}
\end{equation}
where for simplicity we have defined:

\begin{equation}
h_D=\frac 1{\sqrt{2}}\frac{h^lk_1+\widetilde{h}^lk_2}{k_{+}}.
\end{equation}

Given the phenomenological condition $\upsilon _L\ll k_1,k_2\ll \upsilon _R$%
, $\nu $ and $N$ are aproximate mass eigenstates with masses:

\begin{eqnarray}
m_N &\simeq &\sqrt{2}f\upsilon _R,  \label{neutrinor} \\
m_\nu &\simeq &\sqrt{2}\left[ f\upsilon _L-\frac{h_D^2k_{+}^2}{2f\upsilon _R}%
\right] .  \label{neutrinol}
\end{eqnarray}

Additionally, the electron mass is given by:

\begin{equation}
m_e=\frac 1{\sqrt{2}}\left( h^lk_2+\widetilde{h}^lk_1\right) =h_D^ek_{+},
\label{electron}
\end{equation}
with:

\begin{equation}
h_D^e=\frac 1{\sqrt{2}}\frac{h^lk_2+\widetilde{h}^lk_1}{k_{+}}.
\end{equation}

Normally, we expect $h_D$ and $h_D^e$ to be similar in size. Then,
substituting eq. (\ref{neutrinor}) and eq. (\ref{electron}) into eq. (\ref
{neutrinol}) and taking into account that $k_{+}^2\approx \upsilon
_L\upsilon _R$, we arrive at the following expression for $\upsilon _R$ in
terms of $k_{+}$ and the masses of $\nu ,N,$ and $e$:

\begin{equation}
\upsilon _R^2\approx k_{+}^2\frac{m_N^2}{m_\nu m_N+m_e^2}.
\end{equation}

We can see from this expression that the minimum value that $\upsilon _R$
can take is determined by the lower bound on the mass of $N$ and the upper
bound on the mass of $\nu $.

Taking the following central values from \cite{pdg}:

\begin{eqnarray}
m_e &=&5.11\times 10^{-4}\textbf{ }GeV,  \nonumber \\
m_\nu &<&3\times 10^{-9}\textbf{ }GeV,  \nonumber \\
m_N &>&73.5\textbf{ }GeV,
\end{eqnarray}
we arrive at the lower bound for $\upsilon _R$:

\begin{equation}
\upsilon _R>2.7\times 10^7\textbf{ }GeV.
\end{equation}

\newpage\  

\begin{center}
$
\begin{tabular}{||c|c|c|c|c|c|c|c|c|c||}
\hline\hline
& $\phi _{-}^r$ & $\phi _{+}^r$ & $\delta _R^r$ & $\delta _L^r$ & $\phi
_{-}^i$ & $\phi _{+}^i$ & $\delta _R^i$ & $\delta _L^i$ & $O.M.$ $Masses$ \\  
\hline
$\phi _A^0$ & $0.50$ & $0$ & $0$ & $0$ & $0$ & $0.50$ & $0.71$ & $0$ & $%
\upsilon _R$ \\ \hline
$\phi _B^0$ & $-0.50$ & $0$ & $0$ & $0$ & $0$ & $-0.50$ & $0.71$ & $0$ & $%
\upsilon _R$ \\ \hline
$\phi _C^0$ & $0$ & $0.71$ & $0$ & $0.50$ & $0$ & $0$ & $0$ & $0.50$ & $%
\upsilon _R$ \\ \hline
$\phi _D^0$ & $0$ & $0$ & $0$ & $-0.71$ & $0$ & $0$ & $0$ & $0.71$ & $%
\upsilon _R$ \\ \hline
$\phi _E^0$ & $0$ & $0.71$ & $0$ & $-0.50$ & $0$ & $0$ & $0$ & $-0.50$ & $%
\upsilon _R$ \\ \hline
$\phi _F^0$ & $-0.58$ & $0$ & $0$ & $0$ & $0.58$ & $0.58$ & $0$ & $0$ & $k$
\\ \hline
$\phi _G^0$ & $-0.41$ & $0$ & $0$ & $0$ & $-0.82$ & $0.41$ & $0$ & $0$ & $0$
\\ \hline
$\phi _H^0$ & $0$ & $0$ & $-1.00$ & $0$ & $0$ & $0$ & $0$ & $0$ & $0$ \\  
\hline\hline
\end{tabular}
$
\end{center}

Table $1$. Case $I$: Normalized components of the neutral mass eigenstates
and orders of magnitude for the masses. The most general way to express the
scalar mass eigenstates (first column of the table) is through a linear
combination of the flavour eigenstates (first row of the table). The values
in the table mean the corresponding weight in the linear combination. We
have approximated these values to two significative digits. In the last
column we present the order of magnitude $\left( O.M.\right) $ for the
masses of the eigenstates of the first column. This applies for all the
other tables.

\vspace{1cm}

\begin{center}
\begin{tabular}{||c|c|c|c|c|c||}
\hline\hline
& $\phi _{-}^{+}$ & $\phi _{+}^{+}$ & $\delta _R^{+}$ & $\delta _L^{+}$ & $%
O.M.$ $Masses$ \\ \hline
$\phi _A^{+}$ & $-0.71i$ & $0.71$ & $0$ & $0$ & $\upsilon _R$ \\ \hline
$\phi _B^{+}$ & $0$ & $0$ & $0$ & $1.00$ & $\upsilon _R$ \\ \hline
$\phi _C^{+}$ & $0.71$ & $-0.71i$ & $0$ & $0$ & $0$ \\ \hline
$\phi _D^{+}$ & $0$ & $0$ & $1.00$ & $0$ & $0$ \\ \hline\hline
\end{tabular}
\end{center}

Table $2$. Case $I$: Normalized components of the singly charged mass
eigenstates and orders of magnitude for the masses.

\vspace{1cm}

\begin{center}
\begin{tabular}{||c|c|c|c||}
\hline\hline
& $\delta _R^{++}$ & $\delta _L^{++}$ & $O.M.$ $Masses$ \\ \hline
$\phi _A^{++}$ & $0.71$ & $0.32-0.63i$ & $\upsilon _R$ \\ \hline
$\phi _B^{++}$ & $-0.32-0.63i$ & $0.71$ & $\upsilon _R$ \\ \hline\hline
\end{tabular}
\end{center}

Table $3$. Case $I$: Normalized components of the doubly charged mass
eigenstates and orders of magnitude for the masses.

\newpage\  

\begin{center}
$
\begin{tabular}{||c|c|c|c|c|c|c|c|c|c||}
\hline\hline
& $\phi _{-}^r$ & $\phi _{+}^r$ & $\delta _R^r$ & $\delta _L^r$ & $\phi
_{-}^i$ & $\phi _{+}^i$ & $\delta _R^i$ & $\delta _L^i$ & $O.M.$ $Masses$ \\  
\hline
$\phi _A^0$ & $0.50$ & $0$ & $0.71$ & $0$ & $0$ & $0.50$ & $0$ & $0$ & $%
\upsilon _R$ \\ \hline
$\phi _B^0$ & $-0.50$ & $0$ & $0.71$ & $0$ & $0$ & $-0.50$ & $0$ & $0$ & $%
\upsilon _R$ \\ \hline
$\phi _C^0$ & $0$ & $1.00$ & $0$ & $0$ & $0$ & $0$ & $0$ & $0$ & $\upsilon
_R $ \\ \hline
$\phi _D^0$ & $0$ & $0$ & $0$ & $-0.81$ & $0$ & $0$ & $0$ & $0.58$ & $%
\upsilon _R$ \\ \hline
$\phi _E^0$ & $0$ & $0$ & $0$ & $-0.58$ & $0$ & $0$ & $0$ & $-0.81$ & $%
\upsilon _R$ \\ \hline
$\phi _F^0$ & $0.41$ & $0$ & $0$ & $0$ & $0.82$ & $-0.41$ & $0$ & $0$ & $k$
\\ \hline
$\phi _G^0$ & $-0.41$ & $0$ & $0$ & $0$ & $0.41$ & $0.41$ & $0.71$ & $0$ & $%
0 $ \\ \hline
$\phi _H^0$ & $-0.41$ & $0$ & $0$ & $0$ & $0.41$ & $0.41$ & $-0.71$ & $0$ & $%
0$ \\ \hline\hline
\end{tabular}
$
\end{center}

Table $4$. Case $II$: Normalized components of the neutral mass eigenstates
and orders of magnitude for the masses.

\vspace{1cm}

\begin{center}
\begin{tabular}{||c|c|c|c|c|c||}
\hline\hline
& $\phi _{-}^{+}$ & $\phi _{+}^{+}$ & $\delta _R^{+}$ & $\delta _L^{+}$ & $%
O.M.$ $Masses$ \\ \hline
$\phi _A^{+}$ & $0.71$ & $0.71i$ & $0$ & $0$ & $\upsilon _R$ \\ \hline
$\phi _B^{+}$ & $0$ & $0$ & $0$ & $1.00$ & $\upsilon _R$ \\ \hline
$\phi _C^{+}$ & $0.71i$ & $0.71$ & $0$ & $0$ & $0$ \\ \hline
$\phi _D^{+}$ & $0$ & $0$ & $1.00$ & $0$ & $0$ \\ \hline\hline
\end{tabular}
\end{center}

Table $5$. Case $II$: Normalized components of the singly charged mass
eigenstates and orders of magnitude for the masses.

\vspace{1cm}

\begin{center}
\begin{tabular}{||c|c|c|c||}
\hline\hline
& $\delta _R^{++}$ & $\delta _L^{++}$ & $O.M.$ $Masses$ \\ \hline
$\phi _A^{++}$ & $1.00$ & $0$ & $\upsilon _R$ \\ \hline
$\phi _B^{++}$ & $0$ & $1.00$ & $\upsilon _R$ \\ \hline\hline
\end{tabular}
\end{center}

Table $6$. Case $II$: Normalized components of the doubly charged mass
eigenstates and orders of magnitude for the masses.

\newpage\  

\begin{center}
$
\begin{tabular}{||c|c|c|c|c|c|c|c|c|c||}
\hline\hline
& $\phi _{-}^r$ & $\phi _{+}^r$ & $\delta _R^r$ & $\delta _L^r$ & $\phi
_{-}^i$ & $\phi _{+}^i$ & $\delta _R^i$ & $\delta _L^i$ & $O.M.$ $Masses$ \\  
\hline
$\phi _A^0$ & $0$ & $0.71$ & $0$ & $0$ & $0$ & $0$ & $0.71$ & $0$ & $%
\upsilon _R$ \\ \hline
$\phi _B^0$ & $0$ & $0$ & $0$ & $0$ & $0$ & $1.00$ & $0$ & $0$ & $\upsilon
_R $ \\ \hline
$\phi _C^0$ & $0$ & $-0.71$ & $0$ & $0$ & $0$ & $0$ & $0.71$ & $0$ & $%
\upsilon _R$ \\ \hline
$\phi _D^0$ & $0.50$ & $0$ & $0$ & $0.63$ & $-0.50$ & $0$ & $0$ & $-0.32$ & $%
k$ \\ \hline
$\phi _E^0$ & $0$ & $0$ & $0$ & $0.45$ & $0$ & $0$ & $0$ & $0.89$ & $k$ \\  
\hline
$\phi _F^0$ & $-0.71$ & $0$ & $0$ & $0$ & $-0.71$ & $0$ & $0$ & $0$ & $k$ \\  
\hline
$\phi _G^0$ & $0.50$ & $0$ & $0$ & $-0.63$ & $-0.50$ & $0$ & $0$ & $0.32$ & $%
0$ \\ \hline
$\phi _H^0$ & $0$ & $0$ & $1.00$ & $0$ & $0$ & $0$ & $0$ & $0$ & $0$ \\  
\hline\hline
\end{tabular}
$
\end{center}

Table $7$. Case $III$: Normalized components of the neutral mass eigenstates
and orders of magnitude for the masses.

\vspace{1cm}

\begin{center}
\begin{tabular}{||c|c|c|c|c|c||}
\hline\hline
& $\phi _{-}^{+}$ & $\phi _{+}^{+}$ & $\delta _R^{+}$ & $\delta _L^{+}$ & $%
O.M.$ $Masses$ \\ \hline
$\phi _A^{+}$ & $1.00$ & $0$ & $0$ & $0$ & $\upsilon _R$ \\ \hline
$\phi _B^{+}$ & $0$ & $0$ & $0.71$ & $0.71$ & $k$ \\ \hline
$\phi _C^{+}$ & $0$ & $1.00$ & $0$ & $0$ & $0$ \\ \hline
$\phi _D^{+}$ & $0$ & $0$ & $-0.51$ & $0.86$ & $0$ \\ \hline\hline
\end{tabular}
\end{center}

Table $8$. Case $III$: Normalized components of the singly charged mass
eigenstates and orders of magnitude for the masses.

\vspace{1cm}

\begin{center}
\begin{tabular}{||c|c|c|c||}
\hline\hline
& $\delta _R^{++}$ & $\delta _L^{++}$ & $O.M.$ $Masses$ \\ \hline
$\phi _A^{++}$ & $1.00$ & $0$ & $\upsilon _R$ \\ \hline
$\phi _B^{++}$ & $0$ & $1.00$ & $k$ \\ \hline\hline
\end{tabular}
\end{center}

Table $9$. Case $III$: Normalized components of the doubly charged mass
eigenstates and orders of magnitude for the masses.

\newpage\  

\begin{center}
$
\begin{tabular}{||c|c|c|c|c|c|c|c|c|c||}
\hline\hline
& $\phi _{-}^r$ & $\phi _{+}^r$ & $\delta _R^r$ & $\delta _L^r$ & $\phi
_{-}^i$ & $\phi _{+}^i$ & $\delta _R^i$ & $\delta _L^i$ & $O.M.$ $Masses$ \\  
\hline
$\phi _A^0$ & $0$ & $0$ & $-1.00$ & $0$ & $0$ & $0$ & $0$ & $0$ & $\upsilon
_R$ \\ \hline
$\phi _B^0$ & $0$ & $0$ & $0$ & $0$ & $0$ & $1.00$ & $0$ & $0$ & $\upsilon
_R $ \\ \hline
$\phi _C^0$ & $0$ & $1.00$ & $0$ & $0$ & $0$ & $0$ & $0$ & $0$ & $\upsilon
_R $ \\ \hline
$\phi _D^0$ & $0$ & $0$ & $0$ & $0$ & $0$ & $0$ & $0$ & $1.00$ & $\upsilon
_R $ \\ \hline
$\phi _E^0$ & $0$ & $0$ & $0$ & $1.00$ & $0$ & $0$ & $0$ & $0$ & $\upsilon
_R $ \\ \hline
$\phi _F^0$ & $1.00$ & $0$ & $0$ & $0$ & $0$ & $0$ & $0$ & $0$ & $k$ \\  
\hline
$\phi _G^0$ & $0$ & $0$ & $0$ & $0$ & $-0.71$ & $0$ & $-0.71$ & $0$ & $0$ \\  
\hline
$\phi _H^0$ & $0$ & $0$ & $0$ & $0$ & $0.71$ & $0$ & $-0.71$ & $0$ & $0$ \\  
\hline\hline
\end{tabular}
$
\end{center}

Table $10$. Case $IV$: Normalized components of the neutral mass eigenstates
and orders of magnitude for the masses.

\vspace{1cm}

\begin{center}
\begin{tabular}{||c|c|c|c|c|c||}
\hline\hline
& $\phi _{-}^{+}$ & $\phi _{+}^{+}$ & $\delta _R^{+}$ & $\delta _L^{+}$ & $%
O.M.$ $Masses$ \\ \hline
$\phi _A^{+}$ & $-1.00$ & $0$ & $0$ & $0$ & $\upsilon _R$ \\ \hline
$\phi _B^{+}$ & $0$ & $0$ & $0$ & $-1.00$ & $\upsilon _R$ \\ \hline
$\phi _C^{+}$ & $0$ & $0.71$ & $0.71$ & $0$ & $0$ \\ \hline
$\phi _D^{+}$ & $0$ & $0.71$ & $-0.71$ & $0$ & $0$ \\ \hline\hline
\end{tabular}
\end{center}

Table $11$. Case $IV$: Normalized components of the singly charged mass
eigenstates and mass orders or magnitude.

\vspace{1cm}

\begin{center}
\begin{tabular}{||c|c|c|c||}
\hline\hline
& $\delta _R^{++}$ & $\delta _L^{++}$ & $O.M.$ $Masses$ \\ \hline
$\phi _A^{++}$ & $1.00$ & $0$ & $\upsilon _R$ \\ \hline
$\phi _B^{++}$ & $0$ & $1.00$ & $\upsilon _R$ \\ \hline\hline
\end{tabular}
\end{center}

Table $12$. Case $IV$: Normalized components of the doubly charged mass
eigenstates and orders of magnitude for the masses.


\begin{thebibliography}{99}
\bibitem{leftright}  J. C. Pati and A. Salam, Phys. Rev. \textbf{D10},
(1974) 275;

R. N. Mohapatra and J. C. Pati, Phys. Rev. \textbf{D11}, (1975) 566; \textit{%
ibid }2558;

G. Senjanovic and R. N. Mohapatra, Phys. Rev. \textbf{D12}, (1975) 1502.

\bibitem{pdg}  Particle Data Group, D. E. Groom et. al., \textit{``Review of
Particle Physics''}, Eur. Phys. J. \textbf{C15}, (2000) 1.

\bibitem{langacker}  P. Langacker and S. Uma Sankar, Phys. Rev. \textbf{D40}%
, (1989) 1569.

\bibitem{barenboim1}  G. Barenboim and J. Bernabeu, Z. Phys. \textbf{C73},
(1997) 321.

\bibitem{barenboim2}  G. Barenboim and N. Rius, Phys. Rev. \textbf{D58},
(1998) 065010.

\bibitem{deshpande}  N. G. Deshpande, J. F. Gunion, B. Kayser, and F.
Olness, Phys. Rev. \textbf{D44}, (1991) 837.

\bibitem{barenboim3}  G. Barenboim, M. Gorbahn, U. Nierste, and M. Raidal,  
Phys. Rev. \textbf{D65}, (2002) 095003.

\bibitem{see-saw}  M. Gell-Man, P. Ramond, and R. Slansky, in \textit{%
Supergravity}, ed. P. van Niewenhuizen and D. Freedman (North-Holland 1979);
Print-80-0576 (CERN);

T. Yanagida, in \textit{Proceedings of the Workshop of the Unified Theory
and the Baryon Number in the Universe}, ed. O. Sawada and A. Sugamoto
(Tsukuba 1979);

R. N. Mohapatra and G. Senjanovic, Phys. Rev. Lett. \textbf{44}, (1980) 912;

R. N. Mohapatra and P. B. Pall, \textit{Massive Neutrinos in Physics and
Astrophysics}, World Scientific Lecture Notes in Physics - Vol. \textbf{41},
World Scientific, (1991);

R. N. Mohapatra, \textit{Unification and Supersymmetry}, Springer-Verlag,
(1992).

\bibitem{khasanov}  O. Khasanov and G. P\'{e}rez, Phys. Rev. \textbf{D65},
(2002) 053007.

\bibitem{weinberg}  S. L. Glashow, Nucl. Phys. \textbf{22}, (1961) 579;

S. Weinberg, Phys. Rev. Lett. \textbf{19}, (1967) 1264;

A. Salam, in: \textit{Elementary particle physics: relativistic groups and
analyticity}, Nobel Symp. No. \textbf{8}, ed. N. Svartholm Almquist and
Wiksell, Stockholm, (1968) 367.

\bibitem{b-l}  A. Davidson, Phys. Rev. \textbf{D20}, (1979) 776;

R. N. Mohapatra and R. E. Marshak, Phys. Lett. \textbf{B91}, (1980) 222.

\bibitem{ckm}  N. Cabibbo, Phys. Rev. Lett. \textbf{10}, (1963) 531;

M. Kobayashi and T. Maskawa, Prog. Theor. Phys. \textbf{49}, (1973) 552;

M. K. Gaillard and B. W. Lee, Phys. Rev. \textbf{D10}, (1974) 897;

T. Inami and C. S. Lim, Prog. Theor. Phys. \textbf{65}, (1981) 297;

E. A. Paschos and U. T\"{u}rke, Phys. Rep. \textbf{178}, (1989) 145;

G. Buchalla, A. J. Buras, and M. K. Harlander, Nucl. Phys. \textbf{B337},
(1990) 313;

I. I. Bigi and A. I. Sanda, $CP$ \textit{Violation}, Cambridge Monographs on
Particle Physics, Nuclear Physics and Cosmology, (2000).

\bibitem{spontaneous}  T. D. Lee, Phys. Rev. \textbf{D8}, (1973) 1226;

G. Senjanovic, Nucl. Phys. \textbf{B153}, (1979) 334;

G. Senjanovic and P. Senjanovic, Phys. Rev. \textbf{D21}, (1980) 3253;

D. Chang, Nucl. Phys. \textbf{B214}, (1983) 435;

H. Harari and M. Leurer, Nucl. Phys. \textbf{B233}, (1984) 221;

G. Ecker and W. Grimus, Nucl. Phys. \textbf{B258}, (1985) 328;

M. Leurer, Nucl. Phys. \textbf{B266}, (1986) 147.

\bibitem{spontaneous2}  J. M. Frere, J. Galand, A. Le Yaouanc, L. Oliver, O.
Pene, and J. C. Raynal, Phys. Rev. \textbf{D46}, (1992) 337;

G. Barenboim, J. Bernabeu, and M. Raidal, Nucl. Phys. \textbf{B478}, (1996)
527.

\bibitem{spontaneous3}  G. Barenboim, J. Bernabeu, and M. Raidal, Nucl.
Phys. \textbf{B511}, (1998) 577.

\bibitem{fcnc}  B. Grzadkowski, Z. Phys. \textbf{C22}, (1984) 361.

\bibitem{fcnc2}  J. F. Gunion, J. Grifols, A. Mendez, B. Kayser, and F.
Olness, Phys. Rev. \textbf{D40}, (1989) 1546.

\bibitem{cplepton}  J. Hisano, \textit{Lepton-Flavor Violation at Future
Lepton Colliders and the Atmospheric Neutrino Oscillation}, Talk given at
34th Rencontres de Moriond: Electroweak Interactions and Unified Theories,
Les Arcs, France, 13-20 Mar 1999, hep-ph/9906312;

Y. Farzan, O. L. G. Peres, and A. Yu. Smirnov, Nucl. Phys. \textbf{B612}
(2001) 59;

Y. Itow et al., \textit{The JHF-Kamioka neutrino project}, hep-ex/0106019;

I. Mocioiu and R. Shrock, JHEP \textbf{0111}, (2001) 050;

W.-L. Guo and Z.-Z. Zing, Phys. Rev. \textbf{D65}, (2002) 073020;

Y. F. Wang, K. Whisnant, Z. Xiong, J. M. Yang, and B.-L. Young,
Phys. Rev. \textbf{D65}, (2002) 073021;

V. Barger, D. Marfatia, and K. Whisnant, Phys. Rev. \textbf{D65},
(2002) 073023.

\bibitem{exp}  G. G. Hanson, \textit{Searches for New Particles},
Proceedings of the XX International Symposium on Lepton and Photon
Interactions at High Energies, Rome, Italy, 23-28 July, 2001, hep-ex/0111058;

The OPAL collaboration, G. Abbiendi et al., Phys. Lett. \textbf{B256},
(2002) 221;

C. Pagliarone and E. Vataga, \textit{In Search for Physics beyond the
Standard Model at Tevatron}, Published on the Proceedings of the XIII
Convegno sulla Fisica al LEP, Rome, April 18-20, 2001, hep-ex/0111064;

M. Battaglia, A. Ferrari, A. Kiiskinen, and T. Maki, \textit{Pair production
of charged Higgs bosons at future linear }$e^{+}e^{-}$\textit{\ colliders},
To appear on the Proceedings of the Snowmass 2001 Summer Study, Snowmass CO
(USA), June-July 2001, hep-ex/0112015;

The DELPHI Collaboration, J. Abdallah et al., Eur. Phys. J. \textbf{C23},
(2002) 409;

The DELPHI Collaboration, J. Abdallah et al., Phys. Lett. \textbf{B525}
(2002) 17;

The ALEPH Collaboration, Phys. Lett. \textbf{B526} (2002) 191;

W. Lohmann, $e^{+}e^{-}$ \textit{Physics at }$LEP$ \textit{and a Future
Linear Collider}, hep-ex/0202007.

\bibitem{newp}  A. Datta and A. Raychaudhuri, Phys. Rev. \textbf{D62} (2000)
055002.
\end{thebibliography}
\end{document}